\documentclass[preprint,12pt]{elsarticle}

\usepackage{amssymb}

\usepackage{float}
\usepackage{multirow}
\usepackage{amsmath}
\usepackage{url}

\newcounter{bla}

\journal{Computer Physics Communications}

\begin{document}

\begin{frontmatter}



\title{DWR-Drag: A new generation software for the Double Wall-Ring Interfacial Shear Rheometer's data analysis}


\author[a,b]{Pablo Sanchez-Puga\corref{author}}
\author[b]{Miguel. A. Rubio}

\cortext[author] {Corresponding author.\\\textit{E-mail address:} sanchez-puga@ill.fr}
\address[a]{Institut Laue-Langevin, 38042, Grenoble (France)}
\address[b]{Departamento de Física Fundamental, Facultad de Ciencias, Universidad Nacional de Educación a Distancia (UNED), 28232, Las Rozas (Spain)}

\begin{abstract}
The double wall-ring (DWR) rotational configuration is nowadays the instrument of choice regarding interfacial shear rheometers (ISR) in rotational configurations. Complex numerical schemes must be used in the analysis of the output data in order to appropriately deal with the coupling between interfacial and bulk fluid flows, and to separate viscous and elastic contribution or the interfacial response. We present a second generation code for analyzing the interfacial shear rheology experimental results of small amplitude oscillatory measurements made with a DWR rotational rheometer. The package presented here improves significantly the accuracy and applicability range of the previous available software packages by implementing: i) a physically motivated iterative scheme based on the probe's equation of motion, ii) an increased user selectable spatial resolution, and iii) a second order approximation for the velocity gradients at the ring surfaces. Moreover, the optimization of the computational effort allows, in many cases, for on-the-fly execution during data acquisition in real experiments.
\end{abstract}


\end{frontmatter}



{\bf PROGRAM SUMMARY/NEW VERSION PROGRAM SUMMARY}

\begin{small}
\noindent
{\em Program Title: ``DWR-Drag''} \\
{\em CPC Library link to program files:} (to be added by Technical Editor) \\
{\em Developer's repository link:} (if available) \\
{\em Code Ocean capsule:} (to be added by Technical Editor)\\
{\em Licensing provisions (please choose one):} GPLv3  \\
{\em Programming language: MATLAB and Python}                                   \\
{\em Supplementary material: An additional document illustrating further software tests regarding flow structure and details about the numerical method used is provided as a Supplementary Material.}

\noindent{\em Nature of problem: How to determine the interfacial dynamic moduli of fluid–fluid interfaces from experimental data has been a challenge in the rheologists community because it requires i) to accurately separate the contributions of the drags exerted by the interface and the adjacent bulk phases, and ii) to accurately separate the viscous and elastic contributions to the interface response. Moreover, in most cases, the velocity profiles at the interface and  the bulk phases are not linear and, consequently, simplifying hypothesis about the interfacial and bulk phases velocity fields are useless.}\\
{\em Solution method: The physical model includes the upper and lower bulk fluid phases, represented as Newtonian fluids (Navier-Stokes equations), the equilibrium of stresses at a viscoelastic interface under shear (Boussinesq-Scriven equation), and the probe's equation of motion. The hydrodynamic problem is solved using a second order centered finite differences scheme. The representation of the drag on the probe is much improved by implementing a selectable spatial resolution based on the ring's cross-section dimension and by a second order representation of the velocity gradient close to the ring's walls. An iterative scheme allows for obtaining the flow configuration that best matches the experimental data (the complex amplitude ratio) and, consequently, the optimal value of the interfacial dynamic moduli or, equivalently, the interfacial complex viscosity.}\\

\end{small}


\section{Introduction}
\label{sec:intro}

Interfacial shear rheometry \cite{Miller2009} is an experimental technique that aims at measuring the mechanical response of an interface system by imposing a shear stress and measuring the shear deformation. In the last two decades, new experimental and data analysis techniques have been proposed in the field of interfacial rheometry \cite{Fuller2012,Guzman2018}, mainly by devising geometries that comparatively amplify the importance of the interfacial contribution by reducing probe size and inertia as much as possible. Most interfacial shear rheology techniques work on flat horizontal interfaces prepared in shallow containers provided with moving barriers (the so-called Pockels-Langmuir troughs).

Interfacial shear rheometers (ISRs hereafter) based in two different types of probe displacement have been proposed: i) systems with rectilinear probe displacement and ii) systems with rotational probe displacement. Among the rectilinear motion class, the magnetic needle interfacial shear rheometer \cite{Brooks1999,Tajuelo2015,Tajuelo2016} is the instrument providing today the widest operability window. However, rotational rheometers remain a popular choice because these systems can take advantage of the high-resolution capabilities of bulk rotational rheometers for torque and angle measurements, making them easily configurable in any rheology laboratory.

Among the rotational ISRs, the so-called Double Wall-Ring (DWR) geometry \cite{Vandebril2010} has become the laboratory standard  because of its low inertia and its minimal area of contact between the probe and the bulk fluid phases that surround the interface. The DWR ISR is most often operated in oscillatory experiments, i.e., an oscillatory stress, or strain, is imposed on the film and the corresponding oscillatory response in strain, or stress, is measured. 

The main experimental observables are, then, the time histories of the total torque applied by the rheometer and the angular displacement of the rotor + probe ensemble (modern rheometers usually provide the torque corrected for the rotor + ring probe inertia, too). Then, the amplitude ratio and the relative phase between both oscillating variables are used to recover the values of the rheological properties of the interfacial films, namely, the frequency dependent complex interfacial modulus $G_s^*(\omega) = G_s^\prime (\omega) + i G_s^{\prime\prime}(\omega)$, where $\omega$ is the angular oscillation frequency. $G_s^\prime(\omega)$ is the so-called \emph{storage modulus} and represents the elastic part of the interface's response, and $G_s^{\prime\prime}(\omega)$ is the so-called \emph{loss modulus}, that represents the viscous part of the interface's response. Alternatively, a frequency dependent complex viscosity may be defined as $\eta_s^* (\omega)= -\frac{iG_s^*(\omega)}{\omega}$.

However, obtaining accurate values of the complex modulus or viscosity out of the experimental data, namely, the evolution of the torque and the angular position is far from trivial because of the unavoidable coupling between the velocity fields at the interface and the bulk fluid phases.

Vandebril \emph{et al.} \cite{Vandebril2010} proposed an elegant numerical scheme to extract the values of $G^*(\omega)$ from the complex amplitude ratio, $AR^*(\omega) = \frac{T_0(\omega)}{\theta_0(\omega)}e^{i\delta(\omega)}$, where $T_0(\omega)$ is the torque oscillation amplitude, $\theta_0(\omega)$ is the angular displacement amplitude, and $\delta(\omega)$ is the relative phase between both oscillatory variables. The scheme was inspired by the treatment that had been already developed for the magnetic needle ISR \cite{Reynaert2008,Verwijlen2011} and involved the solution of the Navier-Stokes equations at the bulk fluid phases together with the Boussinesq-Scriven equation at the interface, and devising an ad-hoc iterative scheme that, upon convergence, yielded the values of $\eta^*(\omega)$. The corresponding computer code (that we will label as \emph{first generation}, G1, hereafter) was made freely available by its authors and can be downloaded from  \url{https://softmat.mat.ethz.ch/opensource.html}.   Similar iterative schemes, based on  the solution of the Navier-Stokes equations supplemented with the Boussinesq-Scriven equation and the probe equation of motion have been developed for the rotational bicone \cite{Tajuelo2018} and micro-button \cite{Sanchez-Puga2021} rheometers, and for the different configurations of the magnetic needle ISR \cite{Reynaert2008,Verwijlen2011,Tajuelo2016, Sanchez-Puga2021}. The collective designation of flow field-based data analysis (FFBDA) schemes has been applied to such data processing procedures.

The Vandebril \emph{et al.} G1 code was an important breakthrough in the analysis of interfacial shear rheometry data obtained in rotational rheometers. However, some aspects appear as susceptible of improvement. First, the ad-hoc iterative scheme does not guarantee that it would produce a physically correct solution because it does not take into account the probe's equation of motion. Here we adopt the alternative proposed in \cite{Sanchez-Puga2021} that stems directly from the probe's equation of motion. To our knowledge there is not any such code available nowadays. 

Second, the occurrence of strong displacement and velocity gradients close to the ring, both at the interface and the bulk fluid phases, requires an extremely accurate calculation of the hydrodynamic fields close to the ring. That calculation was limited in the Vandebril \emph{et al.} scheme by two factors: i) the discretization mesh used, with a small constant number (just 3) of mesh nodes at the ring surface, and ii) the first order finite differences numerical scheme used. 

Here we describe and make freely available a \emph{second generation} (G2, hereafter) FFBDA scheme for the output data of DWR interfacial shear rheometers. This new version solves all the aforementioned issues of the G1 code. Additionally, the G1 code was designed to process inertia-corrected torque data. Instead, the G2 package further includes the possibility of making the torque's inertia correction within the package, thereby enabling advanced users to work with raw torque data. Hence, the G2 package offers enhanced versatility in selecting geometric, physical, and discretization parameters. These improvements allow users to easily extract accurate values of the frequency-dependent complex interfacial dynamic moduli from data obtained using the DWR configuration.

Two versions of the software are offered, in MATLAB and Python, respectively. The software package uses as input data the amplitude ratio and the phase difference between the torque and the rotor angular displacement. Such data constitute the typical output  of any modern rotational rheometer. Moreover, the software can work both with total (raw) torque values or inertia corrected torque values at the user's choice. All the calculations here reported have been made assuming input data are inertia corrected torque values. 

The main novelties of this new software package are: i) the implementation of a physically based iterative procedure, by using the probe's equation of motion, ii) the implementation of a more precise discretization scheme close to the ring's surfaces, with a user selectable mesh resolution able to place many nodes at the ring cross-section faces (typically 40) that renders more accurate values of the hydrodynamic fields gradients close to the ring, iii) the implementation of a second order finite differences calculation of the velocity gradients close to the ring cross-section faces. Moreover, an extensive use of vectorized programming and sparse matrices management tools of both MATLAB and Python platforms has dramatically improved the computation time. In summary, the G2 package provides an efficient means of analysis for rheometric data obtained with any DWR geometric configuration and is designed to be user-friendly.

The paper is organized as follows. In Section \ref{sec:model_DWR} we describe the system's geometry and the hydrodynamic model used to obtain the flow field at the interface and the bulk fluid phases. Next, a section containing a thorough description of the software follows. Checks of the package performance are organized in sections covering, respectively, the results of consistency tests, error propagation (from experimental data to the values of the complex viscosity) performance, and a comparison with the results obtained using the previously existing software package, with a brief account of the improvements obtained. A useful example of a starting script can be found in \ref{section:exampleScript}, and detailed information regarding the numerical implementation of the problem is given in the Supplementary Material.

\section{DWR system and model}
\label{sec:model_DWR}

\subsection{DWR geometry}

\begin{figure}[H]
    \centering
    \includegraphics[width = 0.7\linewidth]{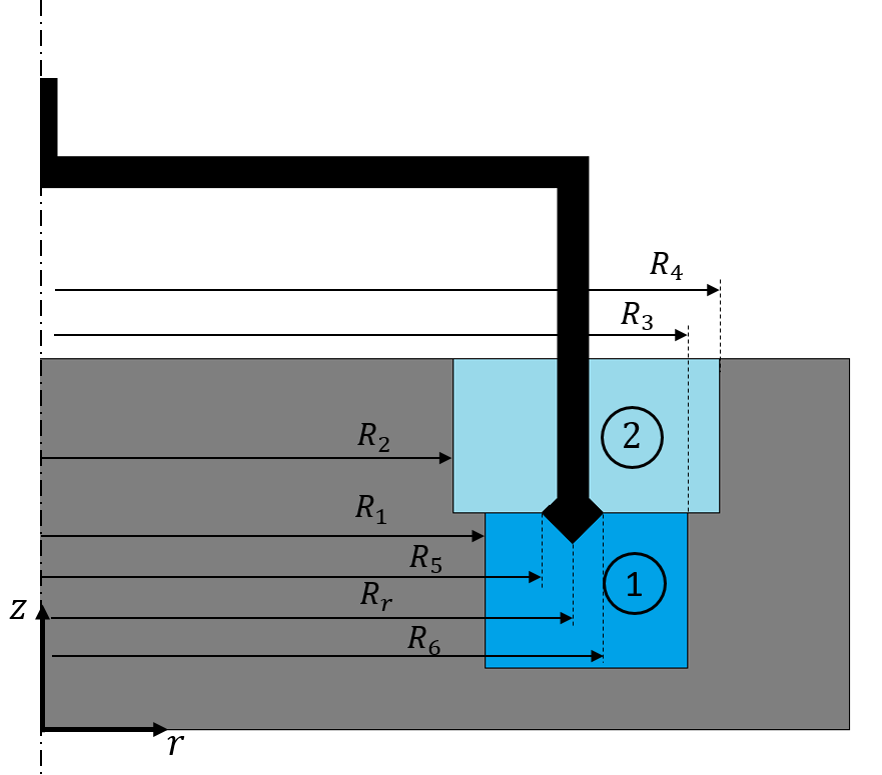}
    \caption{Sketch of the cross-section of the DWR geometry.}
    \label{fig:DWRsketch}
\end{figure}

The geometry of an oscillating DWR system \cite{Vandebril2010} is schematically shown in Figure \ref{fig:DWRsketch}. It consists of a lower annular cylindrical cup, that is filled with the liquid forming the subphase (fluid 1), and a second annular cylindrical cup of slightly larger radial span, that contains the upper fluid bulk phase (fluid 2). The steps separating the upper and lower parts of the cup help to set the interface flat. A thin circular ring with diamond cross section stands on the interface, attached to the rotor of a standard rotational rheometer, with two of its vertices right at the interfacial plane. The ring typically has a few small openings to allow for the inner and outer regions of the interface to be in the same thermodynamic conditions, particularly regarding interfacial surfactant concentration.

\subsection{Physical model}

The physical model used to represent the hydrodynamic flows in the DWR configuration under oscillatory forcing follows the one proposed by Vandebril \emph{et al.} \cite{Vandebril2010}; we reproduce it here briefly for the sake of completeness. The flow field is supposed to be axisymmetric with a single velocity component along the azimuthal direction that depends only on the radial and vertical cylindrical coordinates. The interface is considered to be horizontal and flat, to have zero thickness, and to be confined between the vertices of the DWR and the wall of the shear channel. Gravitational forces, pressure variations, and Marangoni forces are considered negligible. The upper and lower bulk phases are treated as linear viscoelastic fluids, so that the Navier-Stokes equations are valid in each bulk phase. The hydrodynamic drag of the upper bulk phase on the vertical wires that support the ring is not considered.  

The velocity fields at the bulk fluid phases are supposed to oscillate at the forcing frequency, so that they are separable in temporal and spatial contributions; hence, they can be written as \cite{Sanchez-Puga2021}: 

\begin{align}
v_{(1,2)}^*(r,z,t) = \Omega(t)R_6 g^*_{(1,2)}(r,z), 
\label{eq:Ansatz}
\end{align}
where the sub-index (1,2) corresponds to each bulk phase (1, for the subphase; 2, for the upper bulk phase), $\Omega(t)$ is the angular velocity of the rotor, and $g^*_{(1,2)}$ is the non-dimensional complex fluid velocity amplitude function whose real and imaginary components represent the in-phase and out-of-phase components with respect to the probe velocity. Substituting the expression \eqref{eq:Ansatz} in the Navier-Stokes equations, and making the spatial variables dimensionless using the external radius of the ring, $R_6$, where the velocity reaches the maximum value, so that $\Bar{r} = \frac{r}{R_6}$, $\Bar{z} = \frac{z}{R_6}$, the Navier-Stokes equations at the bulk phases read:
\begin{align}
     iRe^*_{(1,2)}g^*_{(1,2)} = \frac{\partial^2 g^*_{(1,2)}}{\partial\Bar{r}^2} + \frac{\partial^2 g^*_{(1,2)}}{\partial\Bar{z}^2} + \frac{1}{\Bar{r}}\frac{\partial g^*_{(1,2)}}{\partial\Bar{r}} - \frac{g^*_{(1,2)}}{\Bar{r}^2},
    \label{eq:NS_DWR}
\end{align}

\noindent where the Reynolds numbers corresponding to each phase are $Re^*_{(1,2)} = \frac{\rho_{(1,2)}\omega R_6^2}{\eta^*_{(1,2)}}$. Here, $\rho_{(1,2)}$ are the mass densities of the bulk phases and $\eta^*_{(1,2)}$ their dynamic viscosities, that are taken as complex quantities to include the possibility of working with viscoelastic subphases. 

No-slip boundary conditions are considered at all contacts with solid walls, including the ring surface, while for the condition at the top of the upper bulk phase the user can select a free interface condition (more appropriate for air/water interfaces) or no-slip in case the upper bulk phase is limited above by a solid container. Detailed expressions for the boundary conditions can be found in the Supplementary Material.

At the interface, the Boussinesq-Scriven equation represents the balance of stresses at the interface, taking into account the coupling with the bulk stresses close to the interface. The boundary condition at the interface, then, reads:

\begin{align}
    \pm N^*\left.\frac{\partial}{\partial\Bar{r}}\left(\frac{1}{\Bar{r}}\frac{\partial}{\partial\Bar{r}}\left(\Bar{r}g^*_{s}\right)\right)\right|_{\Bar{R_{r}}<\Bar{r}<1,\Bar{z}=\Bar{h}}=\left.\frac{\partial g^*_{1}}{\partial\Bar{z}}\right|_{\Bar{R_{r}}<\Bar{r}<1,\Bar{z}=\Bar{h}} - \frac{1}{Y^*}\left.\frac{\partial g^*_{2}}{\partial\Bar{z}}\right|_{\Bar{R_{r}}<\Bar{r}<1,\Bar{z}=\Bar{h}},
    \label{eq:BC_DWR_BS}
\end{align}

\noindent where the $\pm$ symbol refers to the inner and outer contours of the DWR, respectively, and the parameters $N^*$ and $Y^*$ derive from the nondimensionalization process and their expressions are 

\begin{align}
    N^*=\frac{\eta^*_{s}}{R_6\eta^*_{1}};\qquad Y^*=\frac{\eta^*_{1}}{\eta^*_{2}},
\end{align}

\noindent where $N^*$ represents the ratio between interfacial and lower bulk phase viscous shear stresses, and $Y$ represents the ratio between lower and upper bulk phases viscous shear stresses. Hence, to completely specify the dynamical conditions one needs to set the values of four non-dimensional parameters: $Re^*_1$, $Re^*_2$, $N^*$, and $Y^*$. For air/water interfaces $Re^*_2$ and $1/Y^*$ can be safely assumed to be null in most cases. Notice that here $\eta^*_1$ is taken as the reference bulk viscosity for the definitions of $N^*$ and $Y^*$ because such situation is better adapted to water/air interfaces. An equivalent formulation might be done taking $\eta^*_2$ as reference bulk viscosity in the case that the bulk liquid phase 2 has higher viscosity than the bulk liquid phase 1. However, the factor $1/Y^*$ would appear multiplying the vertical gradient of $g^*_1$ in Equation \ref{eq:BC_DWR_BS}, where now $Y^* = \eta^*_2/\eta^*_1$. 

\subsection{Interfacial and bulk stresses: The Boussinesq number}

The non-dimensional parameter $N^*$ is the ratio between the interfacial stress (left hand side of Equation \eqref{eq:BC_DWR_BS}) and the stresses induced at the interface by the bulk subphase (first term of the right hand side of Equation \eqref{eq:BC_DWR_BS}). In interfacial shear rheometry the ratio between interfacial and bulk stresses is called the Boussinesq number, $Bq^*$.  High values of $Bq^*$ indicate that the system is dominated by interfacial stresses; the opposite is true for low values of $Bq^*$. $N^*$ is a possible representation of $Bq^*$ but, in more general terms, and considering only a bulk fluid phase (subphase), $Bq^*$ is typically defined as \cite{Edwards1991}:
\begin{align}
    Bq^* = \frac{\eta_s^*\frac{V}{L_s}P_s}{\eta\frac{V}{L_b}A_b} ,
    \label{eq:Boussi}
\end{align}
\noindent where $\eta_s^*$ is the complex interfacial viscosity, $V$ is the characteristic velocity, $\eta$ is the subphase viscosity, $L_s$ and $L_b$ are characteristic lengths scales for the decay of linear momentum at the interface and in the bulk fluid, respectively. $P_s$ is the perimeter of the contact line at the probe surface, $A_b$ is the area of contact between the probe and the bulk subphase, and $a$ is a lenght scale defined by the probe's area to perimeter ratio, $a = A_b/P_s$. 
Assuming, as in \cite{Vandebril2010}, that velocity gradients at the interface and the bulk phases are similar, one obtains the widely used expression 
\begin{align}
    Bq^* = \frac{\eta_s^*}{\eta a}.
    \label{eq:Boussi2}
\end{align}

For instance, in the case of a ring whose cross-section is a diamond with side $L$ (diagonal of length $\sqrt{2}L$), the $a$ length scale is $a = L$ when only the fluid subphase is considered, and $a = 2L$ when both lower and upper bulk fluid phases are taken into account. Indeed, Vandebril {\it et al.} \cite{Vandebril2010} used expression \ref{eq:Boussi2} to label their results. Hence, to ease the comparison of our results with those of the G1 scheme, in what follows we will use for $Bq^*$ the expression stemming from the physical model, i.e., $Bq^* = N^*$, and $Bq^*_{G1}$ for the values obtained according to the Boussinesq number definition used in \cite{Vandebril2010}.  

Nevertheless, the characteristic lengths $L_b$ and $L_s$ play a very important role regarding the flow configuration. If $L_b$ is smaller than the depth of the subphase the flow profile at the subphase will be nonlinear. Similarly, if $L_s$ is smaller than the radial span of the interface, the flow profile at the interface will be nonlinear too. Hence, interpretations of the experimental results based on the assumption of linear velocity profiles at the bulk phases or the interface will necessarily fail except in the limit of high values of $L_b$ and $L_s$, respectively.

Expressions for the frequency dependent characteristic lengths, $L_s$ and $L_b$, for oscillatory flows have been given by Fitzgibbon \textit{et al.} \cite{Fitzgibbon2014}. 
At the bulk phases, the relevant length scale is  the Stokes viscous length, namely, 
\begin{align}
L_{b} = \left( \frac{\nu}{\omega} \right) ^{1/2}, 
\label{eq:Stokes_length}
\end{align}
\noindent where $\nu = \frac{\eta}{\rho}$ is the kinematic viscosity of the bulk phase and $\rho$ its mass density. To obtain the characteristic length $L_s$, a frequency dependent Boussinesq number is defined, that for air-water interfaces reads \cite{Fitzgibbon2014,Tajuelo2018}
\begin{align}
    &Bq_\omega = \frac{\eta_s}{\eta L_b} = \frac{\eta_s}{\eta} \sqrt{\frac{\rho \omega}{\eta}} ,
    \label{Bq_w}
\end{align}
\noindent and then \cite{Tajuelo2018}
\begin{align}
    L_s = \frac{\eta_s}{\eta \sqrt{Bq_\omega}} = \sqrt{\frac{\eta_s}{\sqrt{\eta\rho\omega}}}
    \label{L_s}.
\end{align}

\subsection{Probe's equation of motion and iterative process}

In order to have a complete representation of the system's dynamics, the probe's (rotor + ring ensemble) equation of motion has to be considered. Assuming that the angular position of the rotor + ring ensemble, $\theta^*(t)$, and the torque imposed by the rheometer on the rotor + probe ensemble, $M^*(t)$, can be written as $\theta^*(t) = \theta_0 e^{i\omega t}$ and $M^*(t) = M_0 e^{i(\omega t- \delta)}$, where $\delta$ is the phase lag of the torque with respect to the angular position, the equation of motion reads

\begin{align}
    M_0 e^{i(\omega t - \delta)} + M^*_1 (r,z,t) + M^*_2 (r,z,t) + M^*_s(r,z,t) = I \ddot{\theta}^*(t),
    \label{eq:Probe_Eq_Mot}
\end{align}

\noindent where $I$ is the moment of inertia of the rotor + ring ensemble, $M^*_1(r,z,t)$ and $M^*_2(r,z,t)$ are the drag torques created by the bulk fluid phases on the probe, and $M^*_s (r,z,t)$ is the torque on the probe exerted by the interface. Detailed expressions for the drag torques are given in the Suppl. Mat.

Dividing Equation \eqref{eq:Probe_Eq_Mot} by the angular position, $\theta_0 e^{i\omega t}$, one can define the time-independent contributions to the complex amplitude ratio between torque and angular position, $AR^*$, which is the main observable:

\begin{align}
    AR^* + AR_1^*((g_1(r,z)) + AR_2^*(g_2(r,z)) + AR_s^*(g_s(r,z)) = -I \omega^2.
    \label{eq:AR_DWR}
\end{align}

Now, if one knows the velocity amplitude functions $g^*_{(1,2)}(r,z)$ and $g^*_s (r)$ and the experimental value of $AR^*$ one may solve Equation \eqref{eq:AR_DWR} for $N^*$, obtaining

\begin{align}
    Bq^* = N^* = \frac{-AR^*_{exp} - I\omega^2 - AR^*_1 (g_1^*) - AR^*_2 (g_2^*)}{i\omega2\pi\eta_1 R_6^3\left(\Bar{R_5}^3\frac{\partial}{\partial\Bar{r}}\left.\left(\frac{g^*_{s}}{\Bar{r}}\right)\right|_{\Bar{r}=\Bar{R_5}} - \frac{\partial}{\partial\Bar{r}}\left.\left(\frac{g^*_{s}}{\Bar{r}}\right)\right|_{\Bar{r}=\Bar{R_6}}\right)},
    \label{eq:Iter_Bous}
\end{align}

\noindent and recover the value of the complex viscosity, $\eta_s^*$. Unfortunately, the amplitude functions $g_{1,2}^*$ and $g_s^*$ depend implicitly on $Bq^*$ because they are the solutions of the Navier-Stokes equations supplemented with the Boussinesq-Scriven boundary condition, where $N^* = Bq^*$ appears.

However, Equation \eqref{eq:AR_DWR} can be solved for $Bq^*$ so, if one knows the numerical solution of problem defined by the Navier-Stokes equations (Equation \eqref{eq:NS_DWR}), with the boundary conditions (Equations (1) to (8)  of the Supp. Mat.), and the Boussinesq-Scriven equation (Equation \eqref{eq:BC_DWR_BS}), together with the experimental value of the complex amplitude ratio, $AR^*_{exp}$, one can obtain the value of $Bq^*$ corresponding to such conditions. This fact can be used to devise an iterative scheme \cite{Sanchez-Puga2021} starting from a \emph{seed} value $Bq_0^*$. The process will involve solving the  Navier-Stokes equations with the aforementioned boundary conditions, obtaining the drag torques, and using the probe's equation of motion (Equation \eqref{eq:AR_DWR}) with the experimental value of the amplitude ratio \cite{Sanchez-Puga2021}, $AR^*_{exp}$,, to obtain an improved value for $Bq^*$, i.e.,
\begin{align}
    [Bq^*]_{k+1} = \frac{-AR^*_{exp} - I\omega^2 - \left[AR^*_1\right]^{k} - \left[AR^*_2\right]^{k}}
    {i\omega2\pi\eta_1 R_6^3\left(\Bar{R_5}^3 \left[\frac{\partial}{\partial\Bar{r}}\left.\left(\frac{g^*_{s}}{\Bar{r}}\right)\right|_{\Bar{r}=\Bar{R_5}}\right]^{k} - \left[\frac{\partial}{\partial\Bar{r}}\left.\left(\frac{g^*_{s}}{\Bar{r}}\right)\right|_{\Bar{r}=\Bar{R_6}}\right]^{k}\right)},
    \label{eq:iteration}
\end{align}

This process can be executed iteratively until a certain level of discrepancy between the experimental and numerical values of $AR^*$, below a tolerance value specified by the user, is met.
The software package here proposed performs according to such a numerical scheme.\\

Mathematically, Equation (\ref{eq:iteration}) configures the data analysis as the problem of obtaining the fixed points of a highly nonlinear iterative map on the complex plane, $[Bq^*]_{k+1} = \mathcal{F} \left([Bq^*]_{k}\right)$. We remark that as all of the equations used here to define the iterative map are physically based, any other iterative map will be acceptable if and only if it shows the same fixed points as the iterative map here defined.

\section{Software description}
\label{sec:soft_desc}

\subsection{Software overview}

The software package is written in the same spirit as previous packages developed by this group to analyse data obtained by means of a torsional rheometer with a bicone probe \cite{Sanchez-Puga2018,Sanchez-Puga2019}. The package reads a data file with CSV format containing experimental values of the angular frequency, $\omega$, the modulus of the amplitude ratio, $|AR^*_{exp}|$, and the corresponding phase difference, $\delta_{exp}$. Then, the package executes the iterative scheme described in the previous Section and outputs the calculated values of the dynamic moduli and other parameters of the calculations in an output file in TXT format.

Full package versions for, both, MATLAB and Python environments are provided. The code has been originally written in the MATLAB (R2023b) environment but we have checked its compatibility with the GNU Octave environment obtaining successful results. The MATLAB (Octave) program version has been tested in OS X and Linux operating systems without any problem.

Alternatively, we have also developed a version of the code written in Python in order to offer another free option to process the data experiments. Python is usually included in modern Linux distributions. Tests with Python in such systems have also been satisfactory.

\subsection{Software package structure}

The software package consists of a main script, where all of the program parameters  are input (see Table \ref{table:Tbl_Param}), and several functions. 
The data analysis is started by means of a script that reads the program parameters and calls the main function \textit{postprocessingDWR\_ll.m} which itself calls two subroutines:

\begin{itemize}
    \item \textit{GetFilenames.m}: this subroutine returns an array of $n$ cells that contains the $n$ experiment filenames at the selected input file path. 
    \item \textit{solve\_NS\_DWR\_ll.m}: this subroutine solves the hydrodynamic equations with the appropriate boundary conditions and returns a column vector, $g_{\alpha}^*$,  containing the values of $g^*(\bar{r},\bar{z})$. Full details on the numerics are given in Section 2 of the Supplementary Material.
\end{itemize}

The dimensionless velocity amplitude functions are then used to compute the hydrodynamic torque contributions in Equation \eqref{eq:AR_DWR}. Then, a new calculated value of the complex amplitude ratio, $[AR^*]^{k+1}$, is obtained and a convergence condition is checked. While convergence is not met the main program repeats the iterative process until the convergence condition is fulfilled.

\subsection{Parameters and Data Input}
\label{sec:Parameters}

We provide the users with a script that facilitates the introduction of the required parameters to execute the main function, ``\textit{postprocessingDWR\_ll}''. An example of the parameter input script is provided with the software package. The input parameters are summarized in five \textit{struct} like variables, namely, Geometry, Mesh, Bulk, iteParams, and IO, that are summarized in Table \ref{table:Tbl_Param}.

\begin{table}[]
\begin{tabular}{|l|l|l|l|}
\hline
Aspect                               & Name           & Units  & Concept                                                                                                                               \\ \hline
\multirow{7}{*}{Geometry}            & \texttt{H}              & m      & Vertical   gap for bulk phases                                                                                                        \\ \cline{2-4} 
                                     & \texttt{R6}             & m      & Ring external radius                                                                                                                   \\ \cline{2-4} 
                                     & \texttt{R5}             & m      & Ring internal radius                                                                                                                    \\ \cline{2-4} 
                                     & \texttt{Rr}             & m      & Ring center radius                                                                                                                     \\ \cline{2-4} 
                                     & \texttt{R1}             & m      & Cup lower internal radius                                                                                                                   \\ \cline{2-4} 
                                     & \texttt{R3}             & m      & Cup lower external radius                                                                                                                   \\ \cline{2-4} 
                                     & \texttt{inertia}        & kg.m2  & \begin{tabular}[c]{@{}l@{}}Moment of inertia of  the\\  rotor + bicone assembly\end{tabular} \\ \cline{2-4} 
                                     & \texttt{ICorrected}     &        &  Torque data are inertia corrected? (Default: `true') \\ \hline
\multirow{3}{*}{Mesh}                & \texttt{ringSubs}       &        & \begin{tabular}[c]{@{}l@{}}Subintervals on the diagonal of the DWR\\ (must be an even number! Default: 40)\end{tabular}                                    \\ \cline{2-4} 
                                     & \texttt{upperBC}        &        & \begin{tabular}[c]{@{}l@{}}Boundary condition at the upper fluid layer\\   (default: 'fb', free boundary)\end{tabular}                    \\ \cline{2-4} 
                                     & \texttt{DOrder}         &        & \begin{tabular}[c]{@{}l@{}}Order of the finite differences approximation\\  for drags calculation (default: 2)\end{tabular}                                           \\ \hline
\multirow{4}{*}{Bulk}                & \texttt{rho\_bulk1}     & kg.m-3 & Density of the lower phase                                                                                                            \\ \cline{2-4} 
                                     & \texttt{eta\_bulk1}     & Pa.s   & Viscosity of the   lower phase (possibly complex)                                                                                                       \\ \cline{2-4} 
                                     & \texttt{rho\_bulk2}     & kg.m-3 & Density of the upper   phase                                                                                                          \\ \cline{2-4} 
                                     & \texttt{eta\_bulk2}     & Pa.s   & Viscosity of the   upper phase (possibly complex)                                                                                                        \\ \hline
\multirow{2}{*}{iteParams}           & \texttt{iteMax}         &        & \begin{tabular}[c]{@{}l@{}}Maximum number of iterations allowed\\ (default: 100)\end{tabular}                                                       \\ \cline{2-4} 
                                     & \texttt{tolMin}         &        & Convergence tolerance (default: $10^{-5}$)                                                                                                                \\ \hline
\multirow{5}{*}{IO} & \texttt{colIndexFreq}   &        & \begin{tabular}[c]{@{}l@{}}Ordinal number of the data file column\\  that contains the frequency of\\  the waveform (default: $1$)\end{tabular}      \\ \cline{2-4} 
                                     & \texttt{colIndexAR}     &        & \begin{tabular}[c]{@{}l@{}}Ordinal number of the data file column\\ that contains the modulus of the\\ amplitude ratio (default: $2$)\end{tabular}   \\ \cline{2-4} 
                                     & \texttt{colIndexDelta}  &        & \begin{tabular}[c]{@{}l@{}}Ordinal   number of the data file column\\  that contains the phase of\\  the amplitude ratio (default: $3$)\end{tabular} \\ \cline{2-4} 
                                     & \texttt{inputFilepath}  &        & Path   to the Input File                                                                                                              \\ \cline{2-4} 
                                     & \texttt{outputFilepath} &        & Path   to the Output File                                                                                                             \\ \hline
\end{tabular}
\caption{Input parameters of the software package to be set by the user. International System units are indicated for all dimensional parameters.}
\label{table:Tbl_Param}
\end{table}

Among the parameters required by the program, there are some geometrical and dynamical parameters of the rheometer, and the physical parameters of the bulk fluid phases. All of them are defined in Table \ref{table:Tbl_Param} although some of them deserve particular comments.

In agreement with our previous remark, all along the program, the subphase viscosity is a complex variable. Hence, viscoelastic subphases can be accounted for by just inserting the value (must be known in advance) of the complex bulk subphase viscosity corresponding to the frequency and temperature at which the interfacial rheology measurements were made.

The parameters located in the Mesh group at Table \ref{table:Tbl_Param} are particularly relevant. The number of mesh subintervals in the diagonal of the diamond cross-section of the ring (\texttt{ringSubs}), controls the spatial resolution of the computation and by construction must be an even number. The parameter \texttt{upperBC} is used to communicate to the package whether the boundary condition at the top of the upper bulk fluid layer is a free boundary (\texttt{upperBC} = `fb') or a no slip condition (\texttt{upperBC} = `ns'). The parameter \texttt{DOrder} sets the type of approximation used for the calculation of the spatial gradients (acceptable values are 1 for the first order calculation and 2 for the second order calculation). 

The convergence tolerance (\texttt{tolMin}) is the parameter used to decide whether convergence has occurred, while the maximum number of iterations allowed (\texttt{iteMax}) actually sets an upper limit to the iteration process to preclude the system iterating endlessly in case convergence is not achieved. An example of a typical script is shown in \ref{section:exampleScript}. 

The program is configured to access experimental data, i.e., frequency (in Hz), modulus (nondimensional) and phase (in rad) of the complex amplitude ratio, by reading from a spreadsheet file. By default, the package expects the experimental data files having a name ending by ``\_exp.txt'', and located in a path specified by the user.

The control software of most commercial rotational rheometers directly provides such an output data file which typically has a user selected column structure. The program looks for all files at the current path having a ``\_exp.txt'' end pattern in alphabetical order. Hence, it is mandatory to change the input data file name for it to have such an ending pattern. Then the user must give the program the numbers of the columns that contain, respectively, the frequency at which the measurement was made (\texttt{colIndexFreq}), the amplitude of the torque/angular displacement ratio (\texttt{coIndexAR}), and the phase lag between the angular displacement and the torque (\texttt{colIndexDelta}).   

\subsection{Data Output}
\label{sec:Data_Out}

At the end of the program execution a file is generated, in the specified output path, in which the input file ending pattern ``\_exp.txt'' is replaced with a ``\_out.txt'' ending pattern. The output data file consists of a table containing several columns and where each output data line corresponds to each line of the input data file. Each output line will contain the corresponding values of the frequency, $f$ (in Hz), the interfacial dynamic moduli, $G_s^\prime$ and $G_s^{\prime\prime}$ in N/m, the real and imaginary parts of the complex interfacial viscosity, $\eta_s^\prime$ and $\eta_s^{\prime\prime}$ in Ns/m, the real and imaginary parts of the model's Boussinesq number, $Bq^\prime$ and $Bq^{\prime\prime}$, the real and imaginary parts of the frequency dependent Boussinesq number, $Bq^\prime_\omega$ and $Bq^{\prime\prime}_\omega$, the modulus and argument of the converged torque/angle amplitude ratio, $\lvert AR^*_{calc}\rvert$ and $\arg (AR^*_{calc})$ in Nm/rad, the elapsed time in each iterative process, in seconds, and the number of iterations until convergence. The paths of both, the input and output data files, are set through the script by the user.

For MATLAB users, a \textit{struct} like variable ``results'' can be saved in ``.mat'' that contains:
\begin{itemize}
    \item $f$ (in Hz).
    \item $G_s^*$ as a complex variable array (in N/m).
    \item $[G_s^*]_{linear}$, obtained by linear approximation with and without calibration subtraction as a complex two column array (in N/m).
    \item $\eta_s^*$ as a complex variable array (Ns/m.)
    \item $[\eta_s^*]_{linear}$, obtained by linear approximation with and without calibration subtraction as a complex variable two column array (in Ns/m).
    \item $Bq^*$ as a complex variable array. 
    \item $Bq^*_\omega$ as a complex variable array. 
    \item $|AR^*_{calc}|$ array (in Nm/rad).
    \item $arg(AR^*_{calc})$ array (in rad).
    \item Time elapsed for each line array (in s).
    \item Number of iterations for each line array until convergence.
\end{itemize}

\subsection{General flowchart}

\begin{figure}[H]
\centering
\includegraphics[width=.7\linewidth]{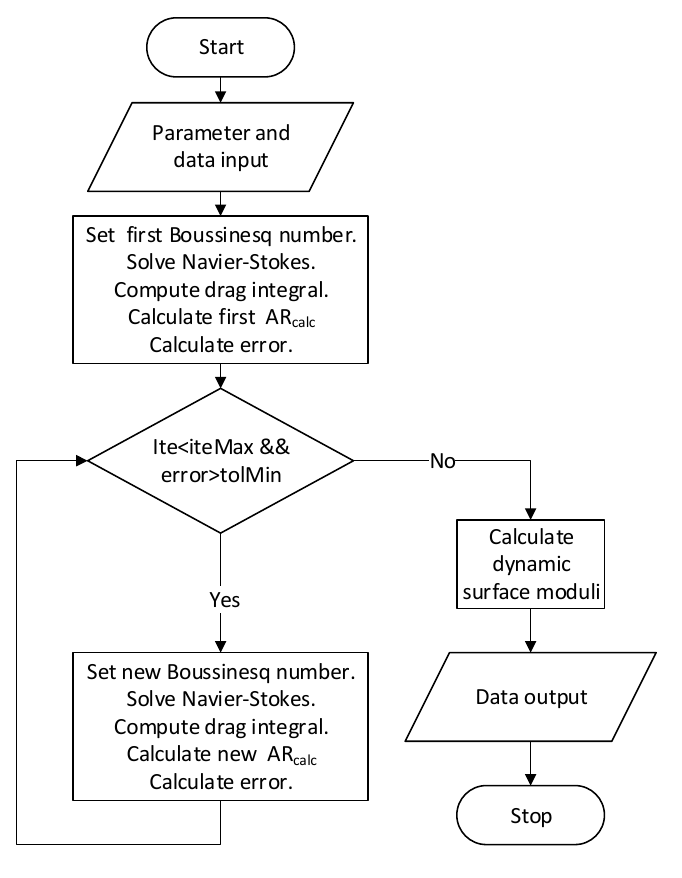}
\caption{General Flowchart}
\label{fig:Flowchart}
\end{figure}

The general flowchart of the program can be seen in Fig. \ref{fig:Flowchart}. First, the required parameter values are input. Then, several tasks are executed in the following order: i) setting a seed value for the model's Boussinesq number $N^*$, ii) solving the Navier-Stokes equation together with the Boussinesq-Scriven equation, iii) calculating the hydrodynamic drag terms, and iv) obtaining the calculated complex amplitude ratio, $AR^*_{calc}$.
Then the value of $AR^*_{calc}$ is checked against $AR^*_{exp}$ for the tolerance value set. If the tolerance criterion is not fulfilled, expression \ref{eq:Iter_Bous} is used to obtain a new approximated value for $N^*$, and steps ii) to iv) are repeated, until convergence is achieved. Whenever the tolerance criterion is fulfilled, the values of the rheological properties are obtained from the converged value of $N^*$ and a new line is written in the output data file.

\subsection{Numerical scheme}
\label{sec:DataAnalysis}

The Navier-Stokes equation, Eq. \eqref{eq:NS_DWR}, with the boundary conditions shown in Equations (2) to (8) of the Supp. Mat., and the Boussinesq-Scriven equation (expression \eqref{eq:BC_DWR_BS}), are solved by means of a second order centered finite differences method. 

\begin{figure}[H]
\centering
\includegraphics[width=.7\linewidth]{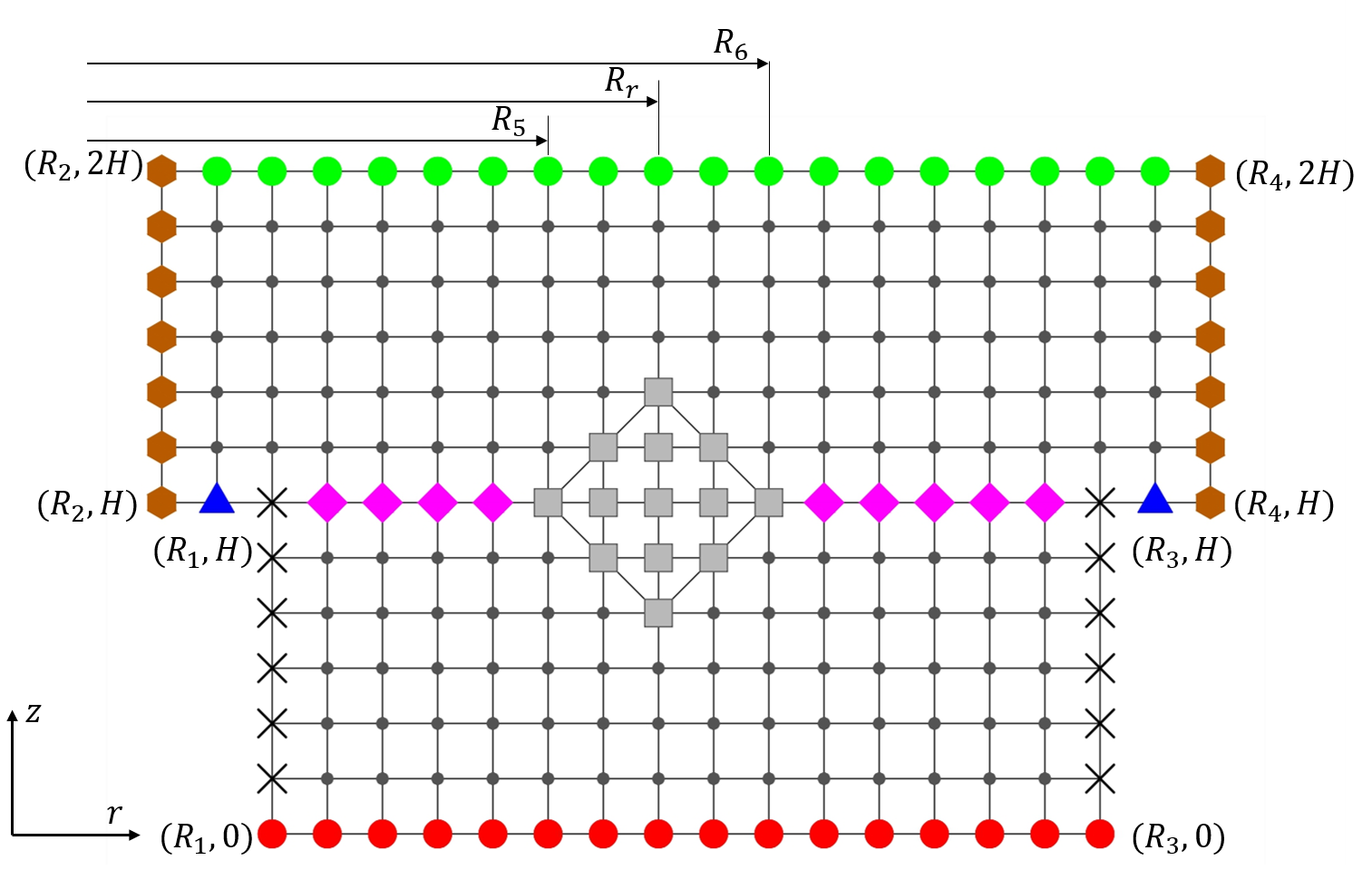}
\caption{Sketch of the mesh and boundaries in a meridian plane with the cylindrical symmetry axis at the left side and the external cup wall at the right side. Red circles: cup's bottom; black crosses: cup's lower lateral wall; blue triangles: cup's half-height horizontal wall; brown hexagons: cup's upper lateral wall. Green circles: top surface of the upper bulk phase; magenta diamonds: interface; grey squares: ring's surface and body. }
\label{fig:Mesh}
\end{figure}

The mesh draws evenly spaced subintervals in the $\bar{r}$ and $\bar{z}$ coordinates, respectively, with the coordinate system origin located at the center of the cup bottom. A cartoon version of the mesh is illustrated in Fig. \ref{fig:Mesh}. Red circles, black crosses, blue triangles, and brown hexagons indicate, respectively, nodes located at the bottom of the cup, the cup's lower lateral wall, the horizontal cup wall at half-height, and the cup's upper lateral wall.   The magenta diamonds represent the nodes located at the interface, while the green circles represent the nodes located at the top surface of the upper bulk phase. The grey squares correspond to the nodes placed at the ring's surface and body. 

In this scheme the ratios between all the lengths that define the cup and ring geometry and the diagonal length of the ring section are kept constant. Then the mesh resolution is set by the number of sub-intervals, $N_d$, located on the diagonal of the ring cross section (an even number by construction) or, equivalently, by the number of sub-intervals on the side of the ring cross section, $N_s$. Both numbers are related by the expression $N_d = 2N_s$. In the software package made available according to \cite{Vandebril2010}, the number of sub-intervals on each side of the ring cross section was fixed to $N_s = 2$, and could not be changed by the user. In the software package described here any even value $N_s \ge 2$ can be used for $N_s$. The limitation to even values for $N_s$ is related to the way that velocity gradients are calculated and it will be explained below. Detailed expressions of the discretization scheme are provided in the Supplementary Material.

Typically $R_6 - R_5 = R_4 - R_3 = 1$ mm and, therefore, $N_d = N_{step}$. In such a mesh and within a centered second order finite differences scheme, the fluid flow equations at each mesh node can be written in terms of the values of $g_{j,k}^*$ at the four nearest neighbors (three for the nodes at the boundaries). 

The integrals appearing in the drag terms at the equation for the complex amplitude ratio (Equation \eqref{eq:AR_DWR}) are calculated by means of the compound trapezoidal rule.

\begin{figure}[H]
\centering
\includegraphics[width=.4\linewidth]{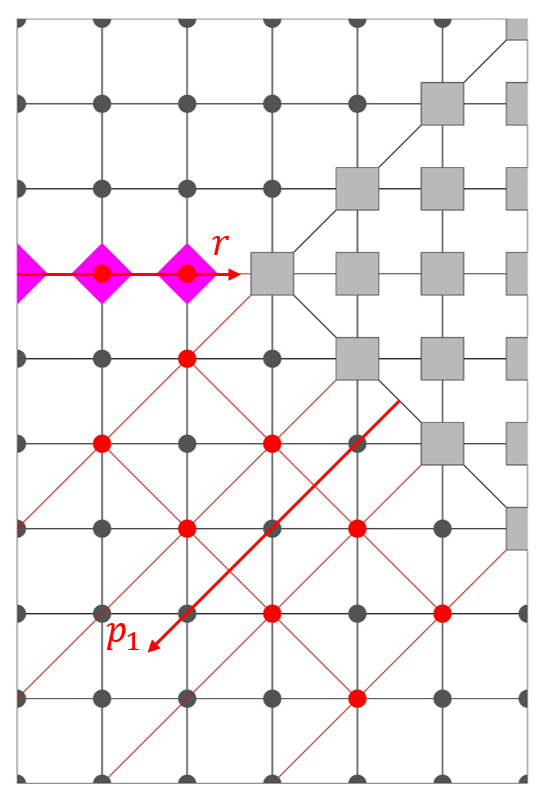}
\caption{Close-up of the mesh near the ring faces. When \texttt{DOrder} = 1, the velocity gradients normal to the ring face are computed using a node at the ring surface and the corresponding first red node along the normal to the ring face. if \texttt{DOrder} = 2, the velocity gradients normal to the ring face are computed using a node at the ring surface and the two corresponding red nodes along the normal to the ring face.}
\label{fig:Mesh_detail}
\end{figure}

A key point here is obtaining accurate values of the velocity gradients close to the ring surface. An accurate calculation of those values is crucial because they are used for the calculation of the hydrodynamic drags that appear in Equation \eqref{eq:AR_DWR}. Actually, the velocity gradients along the normal to the faces of the ring cross-section is the delicate point here. As illustrated in Figure \ref{fig:Mesh_detail}, both the distances between points laying on the ring faces and along the same normal to the ring's cross section is $\Delta p = \sqrt{2}\Delta s$, i.e., the spatial resolution along that direction is worst than along the mesh lines linking nearest neighbour nodes. When the user sets \texttt{DOrder} = 1, a first order finite difference scheme is used and the velocity gradients normal to the ring faces are calculated using only the nodes at the ring surface and those located at the first adjacent line of red nodes parallel to the ring face, as shown in Figure \ref{fig:Mesh_detail}, which has the aforementioned reduced mesh resolution, $\Delta p = \sqrt{2}\Delta s$, i.e.,  
\begin{align}
    {\left(\frac{\partial g^*_{j,k}}{\partial\Bar{p}}\right)}_{i}&=\frac{1}{\Delta\Bar{p}}\left(g^*_{i+1}-g^*_{i}\right).
    \label{eq:Order_1}
\end{align}

If the user sets \texttt{DOrder} = 2, the velocity gradients normal to the ring faces are calculated using a second order difference scheme, that involves also the red nodes located at the second line parallel to the ring faces, as shown in Figure \ref{fig:Mesh_detail}, i.e., 
\begin{align}
    {\left(\frac{\partial g^*_{j,k}}{\partial\Bar{p}}\right)}_{i}&=\frac{1}{2\Delta\Bar{p}}\left(-3g^*_{i} + 4g^*_{i+1} - 2g^*_{i+2}\right).
    \label{eq:Order_2}
\end{align}

Examples of flow field calculations, interfacial velocity profiles and hydrodynamic drag torques under different flow conditions can be found in Section 2 of the Supplementary Material.

\subsection{Iterative process}
\label{Sec:IterativeProcess}

The iterative process is implemented as follows. First, the user must set up  a \emph{seed} value for $N^*$ and, then, an iterative procedure is started till a given convergence criterion is met. At variance with the usual procedure \cite{Vandebril2010,Verwijlen2011,Tajuelo2016,Tajuelo2018}, here we prefer to build the convergence condition not on the value of $N^*$ but on the similarity between the calculated complex amplitude ratio at iteration step $k$, $[AR^*_{calc}]^{k}$ and the experimentally obtained value $AR^*_{exp}$ \cite{Sanchez-Puga2021}, that is the actual output data from the experiment. Nevertheless, comparisons of the results obtained with convergence conditions based either on $N^*$ or $AR^*$ show very minor differences. Consequently, the convergence condition reads

\begin{align}
    \Bigg|\frac{[AR_{calc}^*]^{k}-AR^*_{exp}}{AR^*_{exp}}\Bigg|\leq tolMin,
\label{Eq:tol_condition}
\end{align}

\noindent where $tolMin$ is the convergence threshold set by the user.

The iterative process comprises the following steps:

\begin{enumerate}
\item Solving the Navier-Stokes equation, Eq. \eqref{eq:NS_DWR} , with the boundary conditions given by Eqs. (2) to (8) of the Supp. Mat., and the Boussinesq-Scriven equation \eqref{eq:BC_DWR_BS}, i.e., numerically solving the system of linear equations in expression (6) of the Supplementary Material.
\item Computing the surface and subphase hydrodynamic torques from the solution of the Navier-Stokes equation, and using them to obtain a new value for the calculated complex amplitude ratio, $[AR_{calc}^*]^{k}$, by applying Eq. \eqref{eq:AR_DWR}.
\item Checking whether the iteration scheme has converged or not:
   \begin{itemize}
       \item If the convergence condition is not fulfilled, using the experimental value of the complex torque/angular displacement amplitude ratio and the torque values obtained in step 2 to obtain a new value of the Boussinesq number,  $[N^*]^{k+1}$, by using Equation \eqref{eq:Iter_Bous}. Then repeating steps 1 to 3 till convergence occurs.
       \item If convergence is achieved the iterative process stops.
   \end{itemize}
\end{enumerate}

After reaching convergence, the dynamic surface moduli, $G_s'$ and $G_s''$ are calculated using the definition of the complex Boussinesq number, $N^* = Bq^*$

\begin{align}
G_s^*=-i\omega R_6 \eta_1^* N^*.
\label{G_s}
\end{align}

Generally speaking, the number of iterations required to reach convergence depends on the relative importance of the surface and subphase drags and the rotor inertia: the higher the surface drag, the less iterations are needed. In the simulation tests here reported (see Figure \ref{fig:comp_consistency}) the number of iterations is typically smaller than 20, although for $|Bq^*| \le 100$ it can be as high as 50.

Regarding convergence, in our experience the choice of the \emph{seed} value for $Bq^*$ is not important and good results are usually obtained by choosing as starting value, $[Bq^*]^{k=0} = 0$.
Last, the code makes extensive use of vectorized programming and sparse matrix managing routines from both MATLAB and Python, yielding very fast execution times. 

\section{Program performance}
\label{sec:Performance}

In the following subsections we provide information regarding the tests carried out on different aspects of the software package performance. Specifically, we have checked: i) the influence of the mesh resolution on the accuracy and the computational cost of the hydrodynamic calculations, ii) the effect of the order of approximation in the calculation of velocity gradients, iii) the consistency of the results obtained by means of the iterative scheme, iv) the numerical error propagation from experimental data to the values of the complex viscosity, and v) a brief comparison of results obtained using G1 and G2 software packages. 

Unless otherwise indicated, all results shown in this report have been obtained setting the spatial resolution to $\texttt{ringSubs} = 40$, and using the second order velocity gradient approximation, $\texttt{DOrder} = 2$. The geometrical parameters of the DWR configuration pertain to a medium size DWR system; the complete list of parameter values is given in Table \ref{table:midDWRparams}.

\begin{table}[!ht]
    \resizebox{\textwidth}{!}{%
    \begin{tabular}{|c|c|c|c|c|c|c|c|}
    \hline
        \textbf{H   (mm)} & \textbf{R6   (mm)} & \textbf{R5   (mm)} & \textbf{Rr   (mm) } & \textbf{R1   (mm)} & \textbf{R3   (mm)} & \textbf{stepW   (mm)} & \textbf{I   (Kg.m$^2$)} \\ \hline
        3 & 24.5 & 23.5 & 24 & 20 & 28.78 & 1  & 1.00E-04 \\ \hline
    \end{tabular}%
    }    
    \caption{Geometrical parameters of the medium size DWR.}
    \label{table:midDWRparams}
\end{table}

\subsection{Effects of the mesh resolution}

\begin{figure}[H]
  \begin{minipage}{0.5\textwidth}
  \centering  \includegraphics[width=\linewidth]{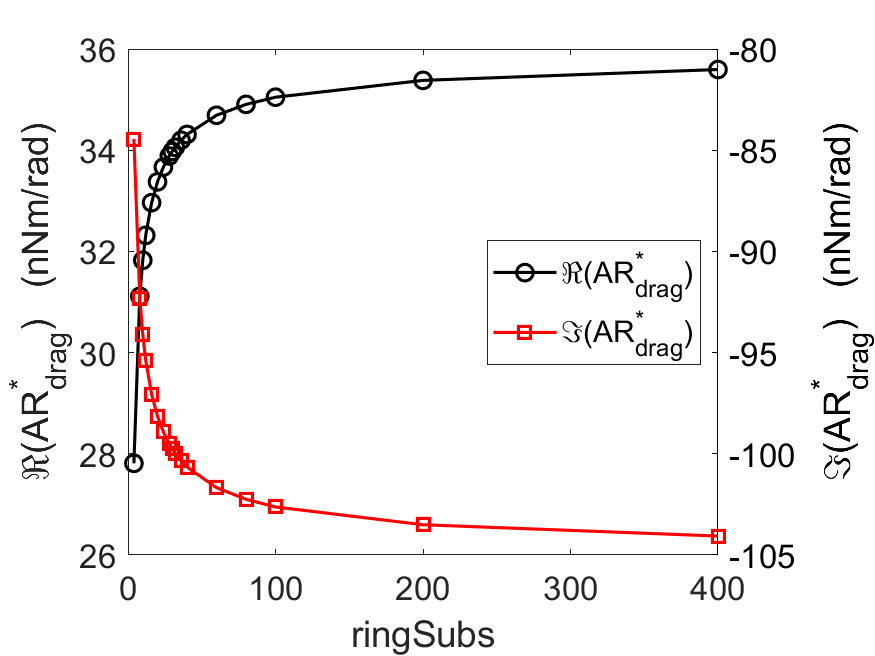}
  \end{minipage}
  \begin{minipage}{0.5\textwidth}
  \centering  \includegraphics[width=\linewidth]{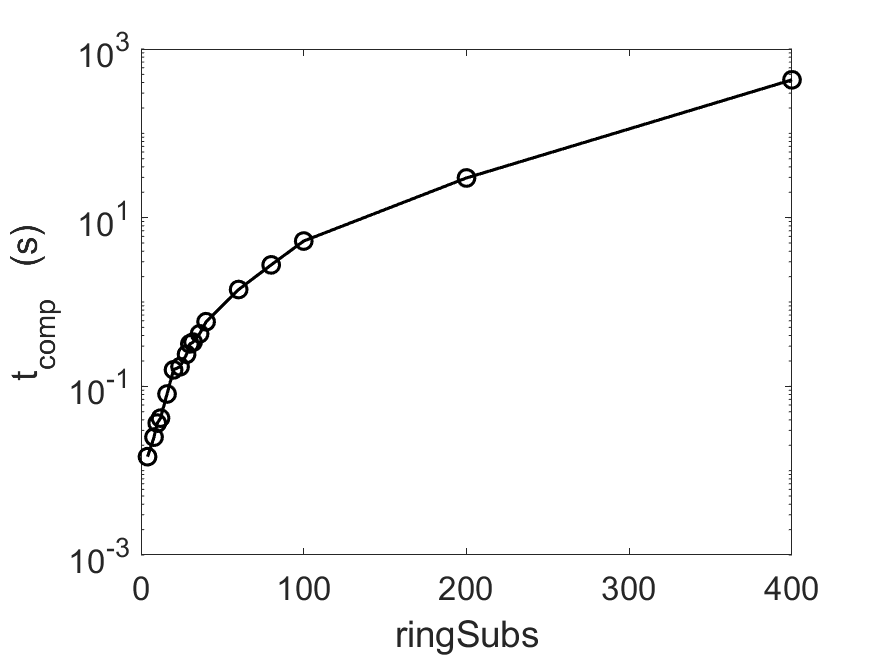}
  \end{minipage}
\caption{Left panel: In-phase (black circles) and out-of-phase (red squares) parts of the amplitude ratio between the total drag torque and the ring displacement for a clean air/water interface ($\eta_s = 0$ Ns/m) at $\omega = 1$ rad/s. Right panel:  time needed to obtain a single flow field configuration in a desktop PC with an 4 nodes Intel Core i5-4590 processor at 3.30 GHz. Note that $t_{comp} \sim 1$ s for $\texttt{ringSubs} = 40$.}
\label{fig:DRW_mesh}
\end{figure}

Increasing the mesh resolution should result in a more accurate calculation of the velocity gradients close to the ring and it should, therefore, affect strongly the values of the different components of the torque on the ring. In Figure \ref{fig:DRW_mesh} (a) and (b) we show, respectively, the dependencies of the in-phase (real) and out-of-phase (imaginary) parts of the amplitude ratio corresponding to the total drag on the ring, $AR^*_{drag} = AR^*_{1} + AR^*_{2} + AR^*_{s}$, and the computer time needed to obtain a single flow field configuration, $t_{comp}$, on the number of mesh nodes at the diagonal of the ring cross section, ($\texttt{ringSubs}$). 

Figure \ref{fig:DRW_mesh}(a) shows the real (black circles curve) and imaginary (red squares curve) parts of the  total torque acting on the ring tend to saturate for values of, approximately, $\texttt{ringSubs} \ge 40$. Indeed, the variations of the in-phase and out-of-phase parts of the total torque are of the order of $5\%$ and $2\%$, respectively, when $\texttt{ringSubs}$ is increased to a value of $400$. 

Figure \ref{fig:DRW_mesh}(b) shows that the computational cost (time needed to obtain a hydrodynamic flow field configuration) is an increasing function of $\texttt{ringSubs}$. However, for $\texttt{ringSubs} = 40$, one gets $t_{comp} \sim 1$ s. Hence, we have used $\texttt{ringSubs} = 40$ throughout the rest of this report because it appears as a good compromise between accuracy and computation time. 

For ease of comparison, whenever the results are shown as a function of the Boussinesq number, the definition by Vandebril \textit{et al.}, i.e., $Bq_{G1} = \frac{\eta_s}{a\eta}$, with $a = 2L = \sqrt{2}$ mm, where $L$ is the side length of the diamond-like ring cross-section, has been used. 

\begin{figure}[H]
  \begin{minipage}{0.5\textwidth}
  \centering  \includegraphics[width=\linewidth]{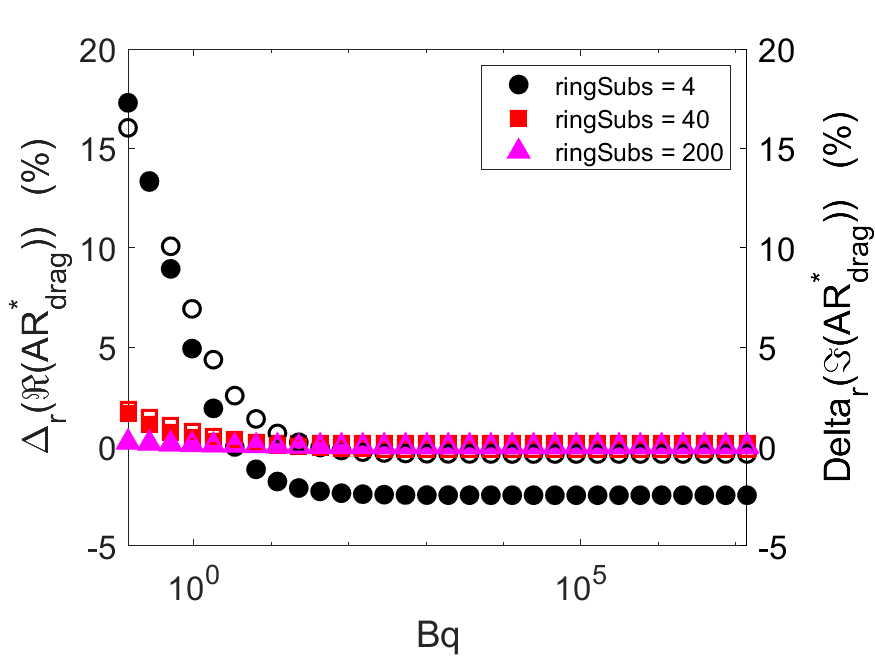}  \end{minipage}
  \begin{minipage}{0.5\textwidth}
  \centering  \includegraphics[width=\linewidth]{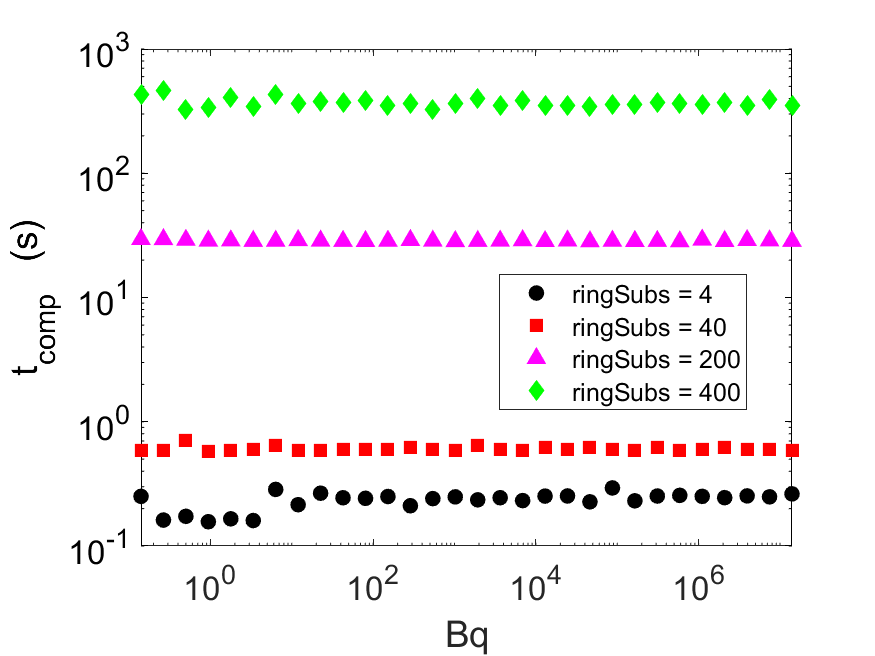}
  \end{minipage}
\caption{Purely viscous interfaces. Left panel: relative error of the in-phase (solid symbols) and out-of-phase (open symbols) components of the amplitude ratio between the total drag torque and the angular displacement at different values of $Bq$ for different mesh sizes. Right panel: $t_{comp}$ dependency on $Bq$ for different values of $\texttt{ringSubs}$.}
\label{fig:DRW_mesh_difrel_signed}
\end{figure}

Is is also interesting to ascertain how much the mesh resolution affects the code performance in a large range of values of $Bq^*$. In the left panel of Figure \ref{fig:DRW_mesh_difrel_signed} we illustrate the error incurred in the calculation of the in-phase and out-of-phase parts of $AR^*_{drag}$ for different mesh resolution, as a function of the Boussinesq number, $Bq_{G1}$, for purely viscous interfaces (notice that in this case $Bq^*$ has no imaginary part). Taking as reference the values obtained for $\texttt{ringSubs} = 400$, we plot in Figure \ref{fig:DRW_mesh_difrel_signed} (left panel) the relative differences of the in-phase (solid symbols) and out-of-phase (open symbols) parts of the total drag torque on the ring for different values of $\texttt{ringSubs}$, namely, $\texttt{ringSubs} = 4,\; 40 \text{, and }200$. Namely, 

\begin{align}
    \Delta_r(\Re(AR^*_{drag})) =  \frac{[\Re(AR^*_{drag})]_{400} - [\Re(AR^*_{drag})]_{\texttt{ringSubs}}}{[\Re(AR^*_{drag})]_{400}} ,\nonumber\\
    \Delta_r(\Im(AR^*_{drag})) = \frac{[\Im(AR^*_{drag})]_{400} - [\Im(AR^*_{drag})]_{\texttt{ringSubs}}}{[\Im(AR^*_{drag})]_{400}}.
\end{align}

For $\texttt{ringSubs} \ge 40$, the relative discrepancies in the real an imaginary parts of the drag torque with the results obtained with $\texttt{ringSubs} = 400$ is well below $1$ $\%$ and $3$ $\%$, respectively. For $Bq \le 10$, both relative errors grow upon decreasing $Bq$, the relative discrepancy being always smaller than $2$ $\%$ in the explored range of $Bq$ provided $\texttt{ringSubs} \ge 40$. We remark that, regarding spatial resolution, the G1 code is equivalent to the case $\texttt{ringSubs}= 4$. Hence, there is a significant improvement in the accuracy in the calculation of the drag components by using the G2 code with $\texttt{ringSubs} \ge 40$.

In the right panel of Figure \ref{fig:DRW_mesh_difrel_signed} we illustrate the corresponding dependency of the computational time, $t_{comp}$, on $Bq$ for meshes with different spatial resolution $\texttt{ringSubs}$. Interestingly, $t_{comp}$ depends strongly on $\texttt{ringSubs}$, but it appears not to depend much on the interfacial viscoelasticity of the film. Hence, for on-the-fly estimations in real-time measurements, $\texttt{ringSubs}\sim 40$ may be a good compromise between accuracy (error in torque in-phase and out-of-phase parts $\le 2$ $\%$) and computational cost ($t_{comp} \le 1$ s) since it yields a computational time comparable to the typical duration of a measurement even in the case that some tens of iterations were needed to achieve convergence. 

\subsection{Effects of the order of the approximation in the calculations of gradients close to the ring}

\begin{figure}[H]
  \begin{minipage}{0.5\textwidth}
  \centering  \includegraphics[width=\linewidth]{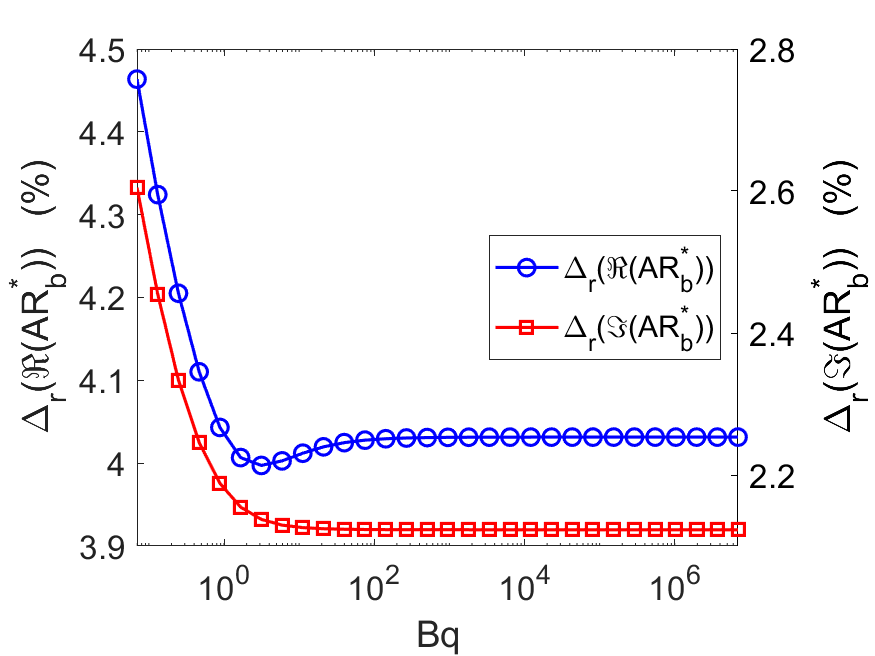}
  \end{minipage}
  \begin{minipage}{0.5\textwidth}
  \centering  \includegraphics[width=\linewidth]{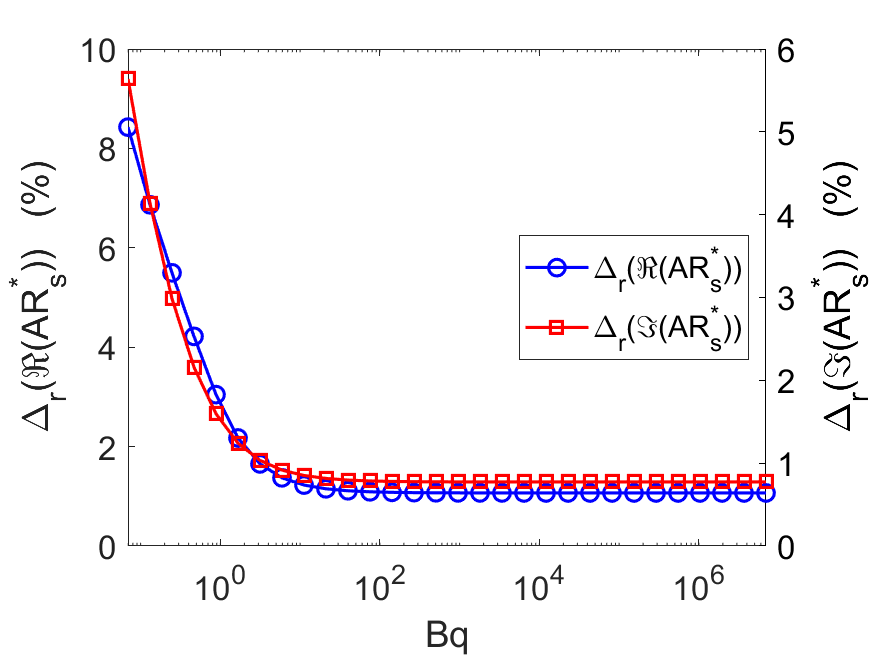}
  \end{minipage}
\caption{Relative differences between the values obtained using $\texttt{DOrder} = 1$ and $\texttt{DOrder} = 2$ for the in-phase (blue circles) and out-of-phase (red squares) parts of the bulk (left panel) and interfacial (right panel) drag torques. Results obtained with $\texttt{ringSubs} = 40$ at $\omega = 1$ rad/s for purely viscous interfaces and shown as a function of $Bq$.}
\label{fig:DRW_DOrder}
\end{figure}

Figure \ref{fig:DRW_DOrder} illustrates the relative differences observed between the results obtained using the first and second order approximations in the case of a purely viscous interface sheared at $\omega  = 1$ rad/s. If we label as $AR^*_s$ the amplitude ratio of the drag exerted by the interface, and $AR^*_b = AR^*_1 + AR^*_2$ the amplitude ratio of the drag exerted by the bulk fluid phases, we define the aforementioned relative differences as

\begin{align}
    \Delta_r(\Re(AR^*_s)) = \left| \frac{[\Re(AR^*_s)]_2 - [\Re(AR^*_s)]_1}{[\Re(AR^*_s)]_2}\right|; \;\;\;\;
    \Delta_r(\Im(AR^*_{s})) = \left| \frac{[\Im(AR^*_{s})]_2 - [\Im(AR^*_{s})]_1}{[\Im(AR^*_{s})]_2}\right|,
\end{align}

\noindent where the sub-index 1 and 2 indicate the order of the gradient approximation, and, conversely, 

\begin{align}
    \Delta_r(\Re(AR^*_b)) = \left| \frac{[\Re(AR^*_b)]_2 - [\Re(AR^*_b)]_1}{[\Re(AR^*_b)]_2}\right|; \;\;\;\;
    \Delta_r(\Im(AR^*_b)) = \left| \frac{[\Im(AR^*_b)]_2 - [\Im(AR^*_b)]_1}{[\Im(AR^*_b)]_2}\right|.
\end{align}

Such relative differences are plotted against $Bq$ in Figures \ref{fig:DRW_DOrder} (left panel) for the bulk phases drag, and \ref{fig:DRW_DOrder} (right panel) for the interfacial drag. 

The graphs in Figure \ref{fig:DRW_DOrder} show that the relative differences between the interfacial and bulk phase's drags obtained with the two gradient approximations increase for $Bq \le 10$ and are rather constant at higher $Bq$ values. For $Bq \ge 10$, the relative differences for the bulk phase's drags are about $4$ $\%$ and $2.2$ $\%$ for the in-phase and out-of-phase parts, respectively. Still in the region $Bq \ge 10$ (see the right panel in Figure \ref{fig:DRW_DOrder}), the interfacial drag torques, the relative differences are for both the in-phase and out-of-phase parts of the interfacial drag torque are below $2$ $\%$ and $2$ $\%$, respectively. In the region $Bq \le 10$, all relative differences increase upon decreasing $Bq$, achieving values above $4$ $\%$ for the bulk drag components and $8$ $\%$ for the interfacial drag components at $Bq \sim 0.1$.  

The computational cost in terms of time, using the first or second order gradient approximation is practically the same, because the operations indicated in the expressions (\ref{eq:Order_1}) and (\ref{eq:Order_2}) bring a very small contribution to the total computational tasks. Consequently, the second order representation, yielding better resolution and accuracy, should be the approximation of choice. The first order approximation has been implemented just to facilitate the comparison with the results obtained using the G1 version (setting \texttt{ringSubs} = 4 and \texttt{DOrder} = 1 mimics the numerical conditions of the G1 version).

\subsubsection{Radial gradient of the interfacial velocity field.}

\begin{figure}[H]
  \begin{minipage}{0.5\textwidth}
  \centering  \includegraphics[width=\linewidth]{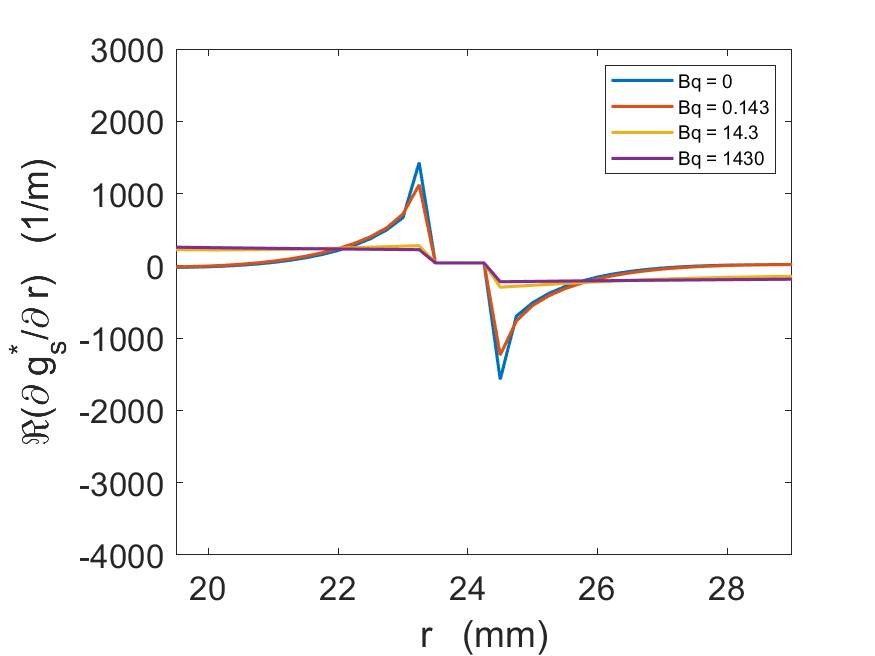}
  \end{minipage}
  \begin{minipage}{0.5\textwidth}
  \centering  \includegraphics[width=\linewidth]{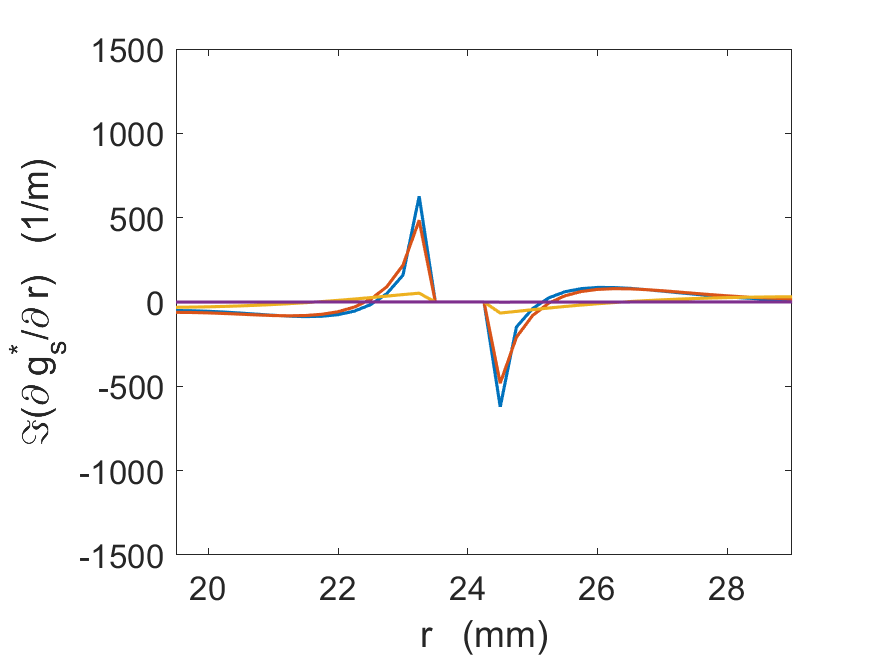}
  \end{minipage}
  \begin{minipage}{0.5\textwidth}
  \centering  \includegraphics[width=\linewidth]{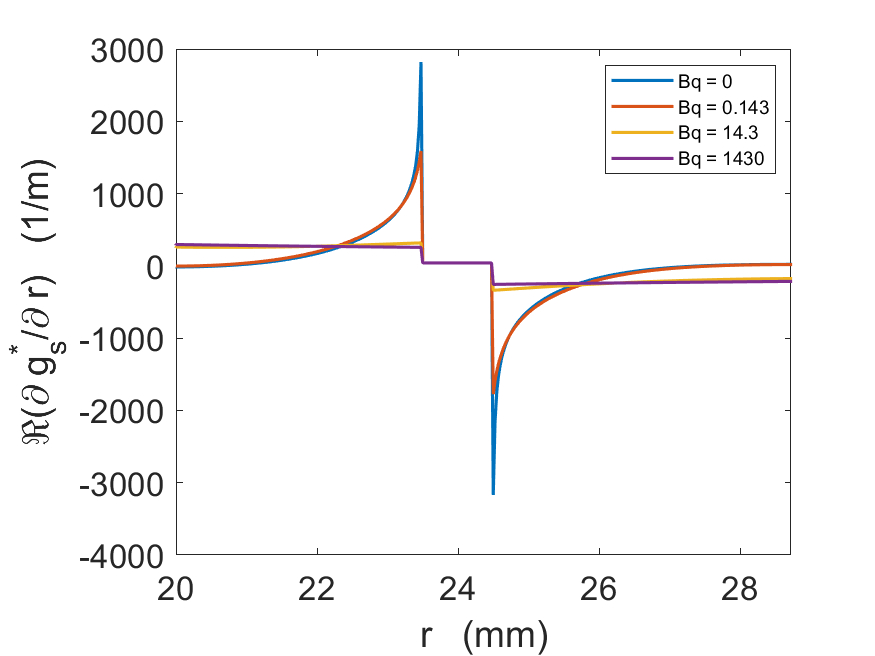}
  \end{minipage}
  \begin{minipage}{0.5\textwidth}
  \centering  \includegraphics[width=\linewidth]{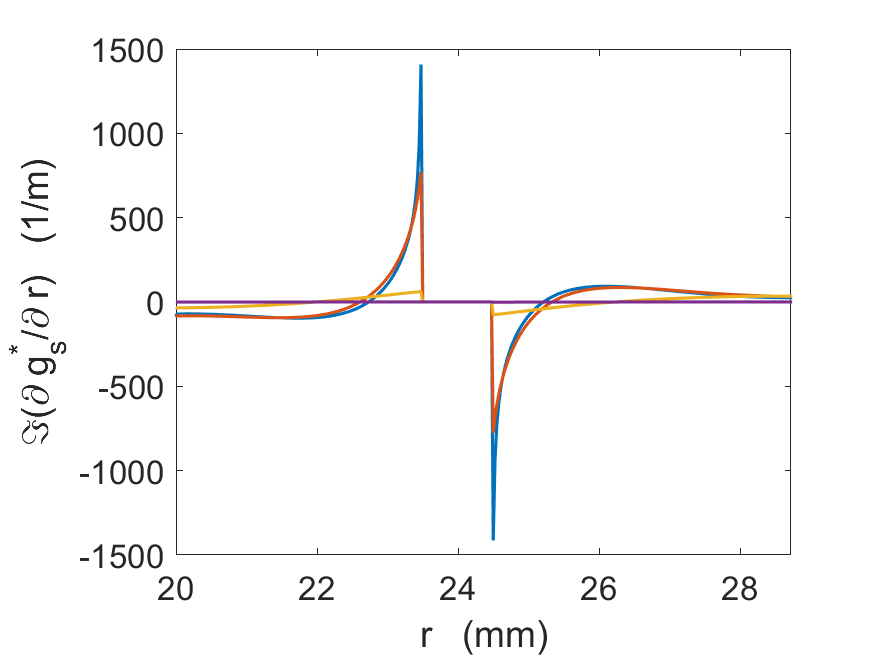}
  \end{minipage}
\caption{Radial gradient of the interfacial velocity amplitude function, $g_s^*(r)$. Panels at left and right: in-phase and out-of-phase parts, respectively for purely viscous air/water interfaces, sheared at a frequency $\omega = 1$ rad/s at different values of $Bq = Bq^*_{G1}$. Upper and lower raw panels were obtained using G1 and G2 code versions, respectively.}
\label{fig:DRW_real_interf_compETH}
\end{figure}

In Figure \ref{fig:DRW_real_interf_compETH} we illustrate the accuracy gain in the calculation of the radial velocity gradients because of the possibility of varying the parameter \texttt{ringSubs} in package G2. In the upper row panels we plot the radial profiles of the in-phase (left panel) and out-of-phase (right panel) parts of the interfacial velocity amplitude function for a purely viscous interface, sheared at $\omega = 1$ rad/s, for different values of $Bq = Bq^*_{G1}$, and obtained with \texttt{ringSubs = 4} (same resolution as package G1). Graphics in the lower panels of Figure \ref{fig:DRW_real_interf_compETH} display the same information but obtained with \texttt{ringSubs = 40} (the recommended value when running the G2 package). 

Regarding the in-phase of the radial velocity gradient, a constant gradient profile is obtained with both packages for the highest value or $Bq$, corresponding to a linear radial  velocity profile, as expected. The differences between both packages arise as $Bq$ diminishes, with a clear piece-wise linear structure appearing for the profiles obtained with the G1 mesh spacing at the lowest values of $Bq$, due to the insufficient spatial resolution of the mesh. Conversely, the gradient profiles obtained with the recommended G2 mesh resolution, shows a much smoother appearance and, more importantly, attains much higher values than those obtained with the G1 mesh resolution, that will result in a much more accurate calculation of the interfacial drag value.
Similar comments regarding the piece-wise structure of the radial gradients may be 
drawn for the imaginary part of the gradients obtained using the mesh resolution of the package G1 (see panels at the right column in Figure \ref{fig:DRW_real_interf_compETH}).

\subsection{Consistency of the iterative procedure}

Consistency tests check that the iterative process does not find spurious solutions \cite{Sanchez-Puga2019,Sanchez-Puga2021}. This is checked by solving the \emph{direct problem}, i.e., setting an initial seed value for the Boussinesq number, of $Bq^*_i$,  and finding the flow field and the corresponding complex amplitude ratio, $AR^*$, and then using that value of $AR^*$ as input for the \emph{inverse problem}, namely, taking $AR^*$ as input to find the corresponding output value $Bq^*_o$ using the iterative scheme. 

\begin{figure}[H]
  \begin{minipage}{0.5\textwidth}
  \centering  \includegraphics[width=\linewidth]{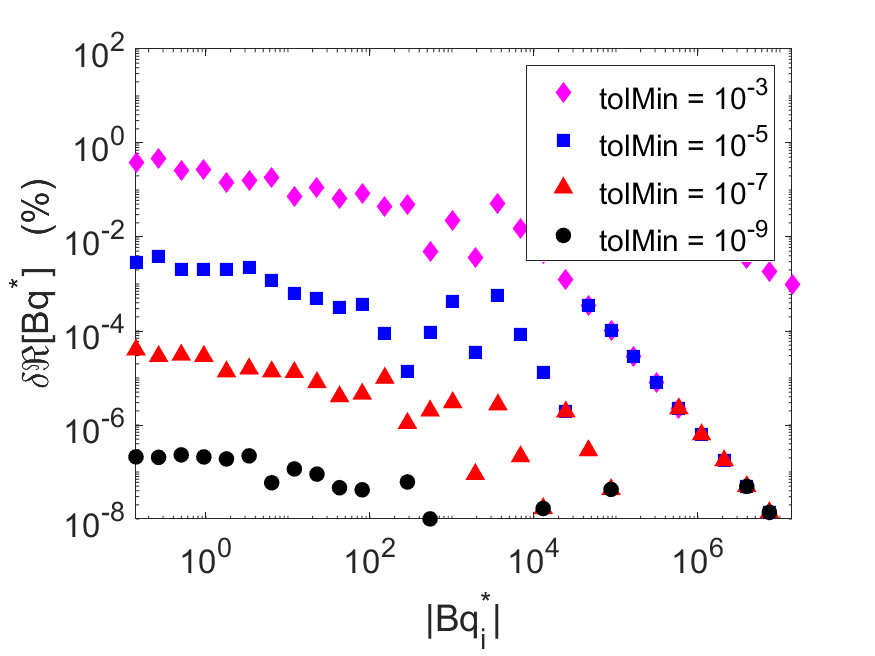}
  \end{minipage}
  \hfill
  \begin{minipage}{0.5\textwidth}
  \centering  \includegraphics[width=\linewidth]{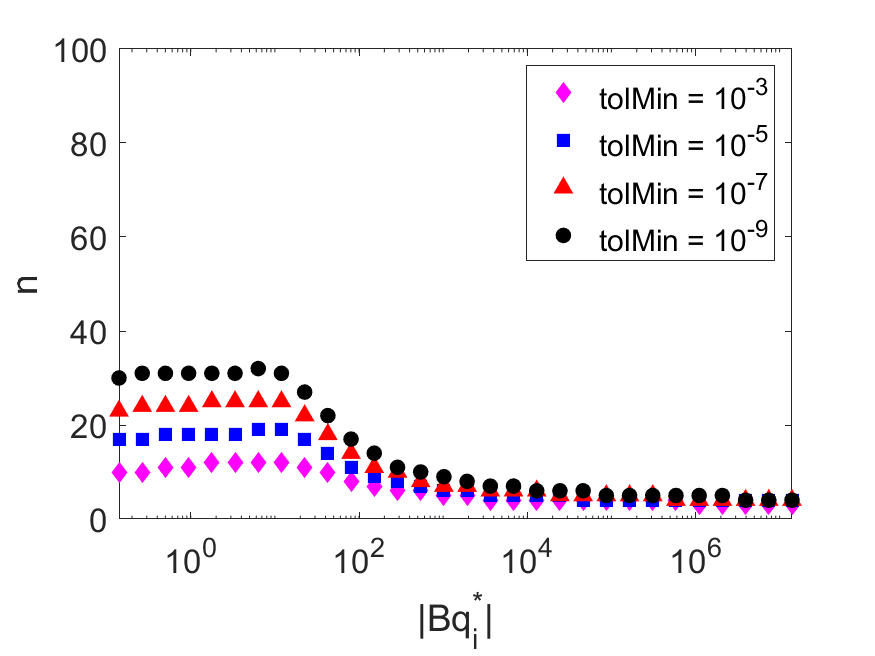}
  \end{minipage}  
  \begin{minipage}{0.5\textwidth}
  \centering  \includegraphics[width=\linewidth]{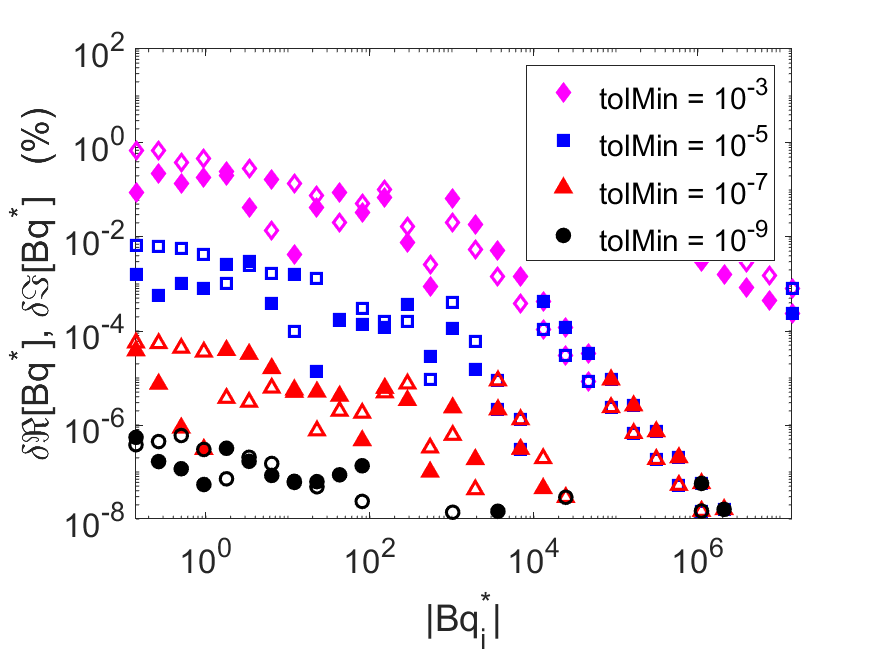}
  \end{minipage}
  \hfill
  \begin{minipage}{0.5\textwidth}
  \centering  \includegraphics[width=\linewidth]{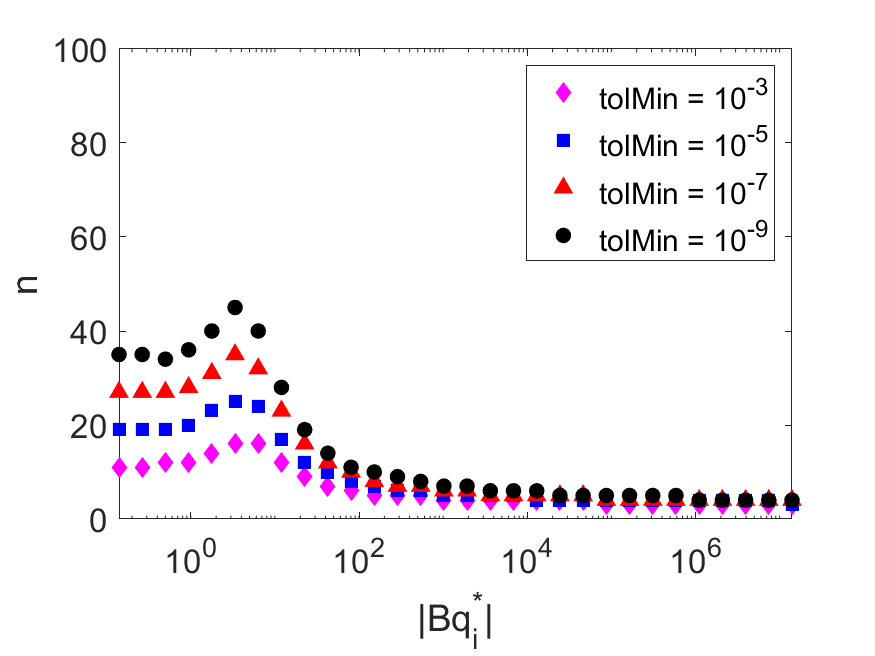}
  \end{minipage}  
  \begin{minipage}{0.5\textwidth}
  \centering  \includegraphics[width=\linewidth]{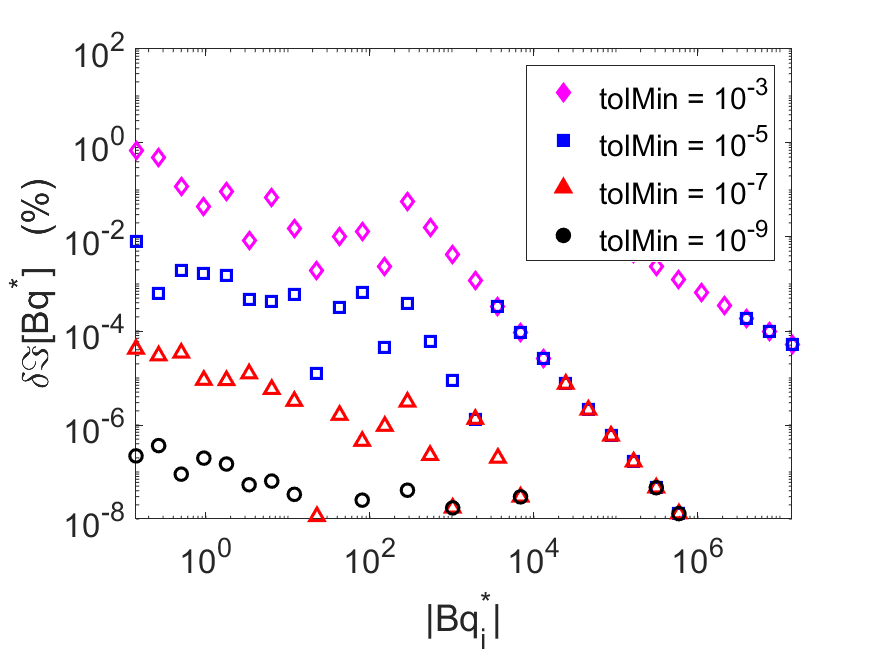}
  \end{minipage}
  \hfill
  \begin{minipage}{0.5\textwidth}
  \centering  \includegraphics[width=\linewidth]{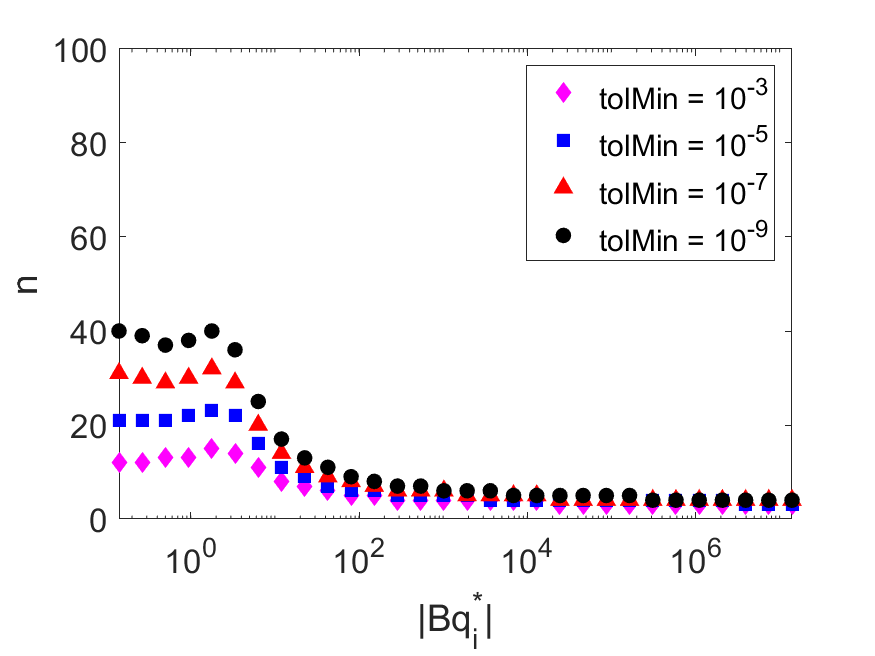}
  \end{minipage}  
\caption{Consistency tests for air/water interfaces. Panels at the left column: relative differences $\delta \Re[Bq^*]$ and $\delta \Im[Bq^*]$ as a function of $|Bq^*_i|$, obtained for different convergence tolerances: $\texttt{tolMin} = 10^{-3}$ (magenta diamonds), $10^{-5}$ (blue squares), $10^{-7}$ (red triangles), and $10^{-9}$ (black circles). Panels at the right column: number of iterations till convergence. Upper row panels: purely viscous interfaces. Middle row panels: viscoelastic interfaces with $\Re[Bq^*] = \Im[Bq^*]$. Lower row panels: purely elastic interfaces.}
\label{fig:comp_consistency}
\end{figure}

We have performed such checks on the G2 code in a wide range of values of $Bq^*_i$ by setting $\omega = 1$ rad/s, and varying the value of $\eta^*_s$ so that the range $10^{-1} \le |Bq^*_i| \le 10^7$ is explored. In Figure \ref{fig:comp_consistency}, we show the results of some illustrative consistency tests. The left column shows the results of the relative differences between the real (loss) and imaginary (storage) parts, respectively, of $Bq^*_i$ and $Bq^*_o$, namely,  $\delta \Re[Bq^*] = \left| \frac{\Re[Bq^*_o] - \Re[Bq^*_i]}{\Re[Bq^*_i]} \right|$ and $\delta \Im[Bq^*] = \left|\frac{\Im[Bq^*_o] - \Im[Bq^*_i]}{\Im[Bq^*_i]}\right|$. The different traces correspond to four different values of the tolerance set in the iterative process, $ \texttt{tolMin} = 10^{-3}$ (magenta diamonds), $10^{-5}$ (blue squares), $10^{-7}$ (red triangles), and $10^{-9}$ (black circles). The right column shows the corresponding values of the number of iterations, $n$, that have been necessary to meet the convergence condition. Three expected general trends can be identified in all of the cases shown in Figure \ref{fig:comp_consistency}: i) the lower the tolerance the smaller the relative differences between input and output values of $Bq^*$, ii) the differences $|Bq_o^* - Bq_i^*|$ are always slightly higher than \texttt{tolMin}, and iii) the lower the tolerance the higher the number of iterations needed to meet the convergence condition. 

The panels in the upper row of Figure \ref{fig:comp_consistency} correspond to the case of purely viscous air/water interfaces with $\eta^*_s = \eta_s$ ($Bq^*$ is a real number). In all cases the relative difference $\delta \Re[Bq^*]$ takes very small values, and it is clear that using $\texttt{tolMin} \le 10^{-5}$ guarantees relative differences smaller than $10^{-2}\,\%$. On the other hand, the number of iterations needed for convergence increases upon decreasing $Bq_i$ and saturates below $Bq_i \sim 10$, showing a plateau for lower values of $Bq_i$. 

The panels in the middle row of Figure \ref{fig:comp_consistency} correspond to the case of viscoelastic air/water interfaces with $\eta^*_s = \eta_s - i \eta_s$ ($Bq^*$ is a complex number). In this case the relative differences are calculated separately for the real (loss) and imaginary (storage) parts of $Bq^*$, open and filled symbols correspond, respectively, to $\delta \Re[Bq^*]$ and $\delta \Im[Bq^*]$. The behavior of the relative differences upon varying $Bq^*_i$ is very similar to the case of the purely viscous interfaces. For these viscoelastic interfaces, however, there is a clear peak in the number of iterations needed for convergence at $|Bq^*_i| \lesssim 10$. Below $|Bq^*_i| \sim 1$ the number of iterations till convergence, $n$, is roughly constant with values very similar to those in the case of purely viscous interfaces.

The panels in the lower row correspond to the case of purely elastic interfaces with $\eta^*_s = -i \eta_s$ ($Bq^*$ is an imaginary number). Again, for these interfaces, both graphs of the relative differences and  the number of iterations needed for convergence are similar to the cases described in the panels corresponding to the top and middle rows.

Therefore, the results of the consistency checks are quite satisfactory because using $\texttt{ringSubs} = 40$ and $\texttt{tolMin} = 10^{-5}$ relative differences below $0.01$ \%, between the programmed and iteratively obtained values of $Bq^*$, can be obtained with computational times below $40$ seconds in the worst case.

\subsection{Error propagation in the iterative process}

Another important aspect of the code's performance is how the code propagates the unavoidable errors in the measurement of the modulus and argument of the complex amplitude ratio, $AR^*$. Here we analyse error propagation by the G2 code by means of the following numerical procedure: i) we select a set of values of $Bq^*_i$, with $\Re[Bq^*_i]$ and $\Im[Bq^*_i]$ taking each 40 equally spaced values in logarithmic scale between $0.14$ and $1.4\times10^7$, ii) for each $Bq^*_i$ we solve the \textit{direct} problem and obtain its corresponding $AR^*$ value, iii) we consider the obtained $AR^*$ value as coming from an experimental measurement that has a relative error of $\pm 1 \%$ both in modulus and argument, iv) we solve the \textit{inverse} problem for the nine points in the perimeter plus the centre of the square defined by $|AR^*| (1+0.01 m)$ and $\delta (1+0.01 n)$, with $m, n = -1, 0, +1$); the largest relative difference between the value of $|AR^*|$ at the eight points in the square perimeter and the value at the square centre is used as an indicator of the error propagated by the iterative process.

\begin{figure}[H]
\centering  \includegraphics[width=.7\linewidth]{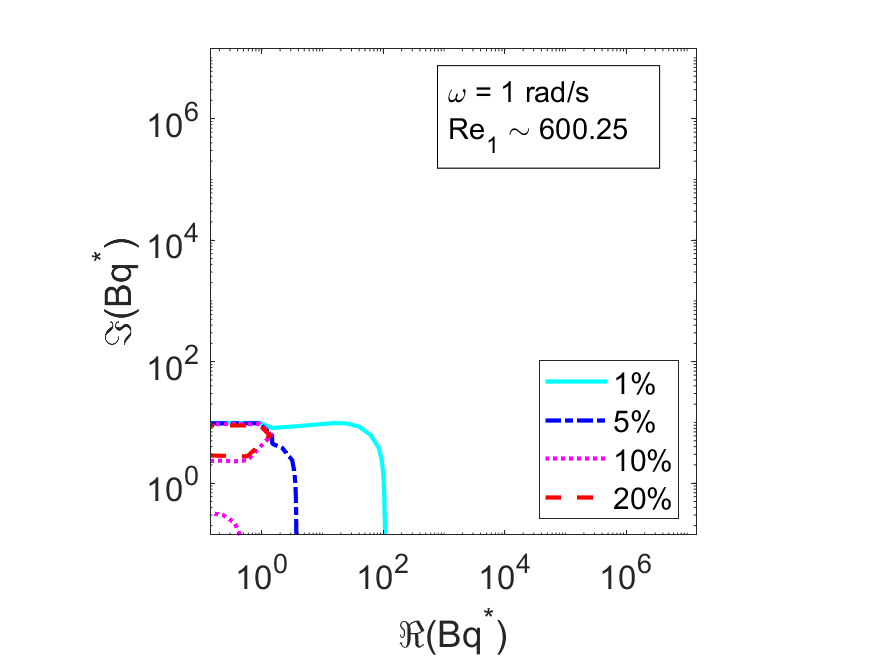}
\caption{Isocontours of the relative error propagated by the G2 code in the $(\Re[Bq^*]$,$\Im[Bq^*])$ plane. The region above and at the right of each isocontour line shows relative errors lower than the label of the line.}
\label{fig:Errors}
\end{figure}

The results of such a study in the $(\Re[Bq^*]$,$\Im[Bq^*])$ plane, using $\texttt{tolMin} = 10^{-5}$, are shown in Figure \ref{fig:Errors}, where isocontour lines of the relative differences (propagated error) calculated are shown. Values of the  propagated error lower than the isocontour line label are found above and at the right of the isocontour line. The results show that the  error propagation due to the iterative scheme of the G2 code is negligible (smaller than $1\%$) for practical purposes in most of the plane, namely, the region delimited by $\Re[Bq^*] \ge 100$ and $\Im[Bq^*] \ge 10$. Propagated errors larger than $5\%$ are found in a very small region for $\Re[Bq^*] \le 1$ and $1 \le \Im[Bq^*] \le 10$.
Notice that here we study the performance of the code in how it propagates the errors in the input data. For a detailed study of how all sources of error affect the operativity window of a particular realisation of a DWR system the interested readers are referred to the paper by Renggli \emph{et al.} \cite{Renggli2020}.

\subsection{Comparative performance with previous methods of analysis}

The next question is to what extent the results of these complicated data analysis schemes are better than the results offered by other simpler data analysis procedures already existing in the literature. 

Operationally, we perform the comparison by, first, setting a nominal input value of the Boussinesq number, $Bq^*_i$, solving the direct problem, and obtaining the corresponding value of the complex amplitude ratio, $AR^*_i$. Second, $AR^*_i$ is given as input to the corresponding data analysis scheme, getting an output value of the Boussinesq number, $Bq^*_o$. 

Here we compare the results of the application of the G2 software package against the results yielded by three other previous data analysis schemes: i) a scheme that disregards the interface-bulk phases coupling and assumes a linear velocity profile at the interface (labelled as scheme L, hereafter), ii) a scheme that assumes a linear velocity profile an a bulk phases drag similar to the drag the bulk phases exert at a clean air/water interface, (labelled as scheme LC, hereafter; see \cite{Sanchez-Puga2021} for a detailed explanation), and iii) the G1 code, that includes the correct interface-bulk phases hydrodynamic coupling but with limited spatial resolution for the velocity gradients representation.

Here, the bulk phases's drags in clean interface conditions are found by solving the direct problem with an interface having $Bq^* = 0$. All of the G2 code results reported in this subsection have been obtained using the recommended values \texttt{tolMin} $= 10^{-5}$, and \texttt{ringSubs} $ = 40$.  

\begin{figure}[H]
  \begin{minipage}{0.5\textwidth}
  \centering  \includegraphics[width=\linewidth]{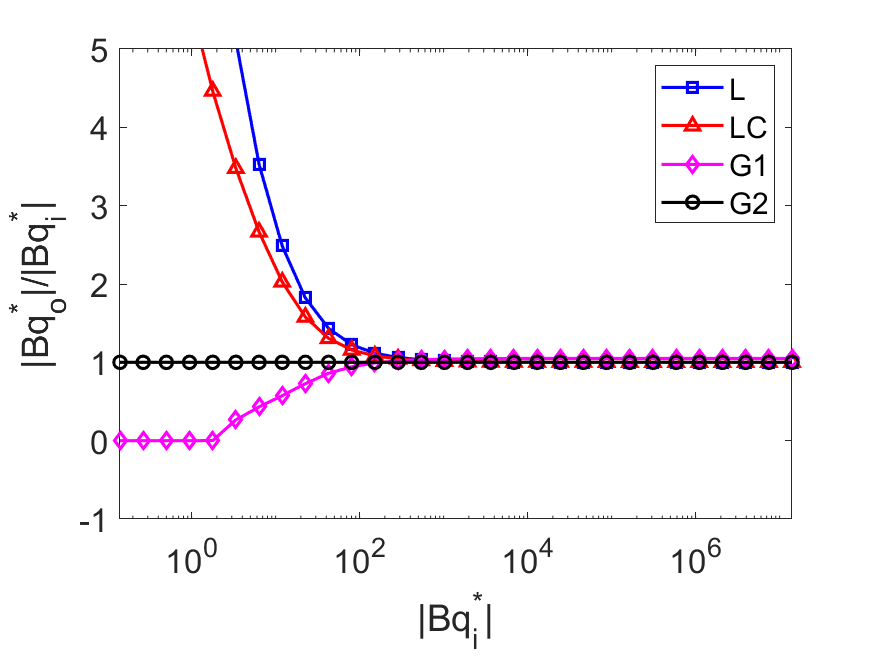}
  \end{minipage}
  \begin{minipage}{0.5\textwidth}
  \centering  \includegraphics[width=\linewidth]{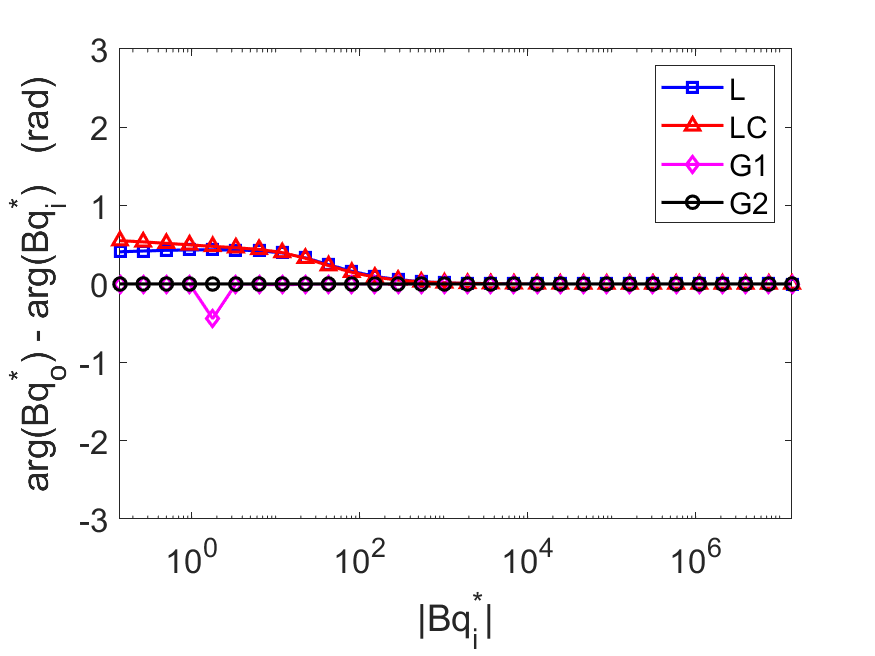}
  \end{minipage}
  \begin{minipage}{0.5\textwidth}
  \centering  \includegraphics[width=\linewidth]{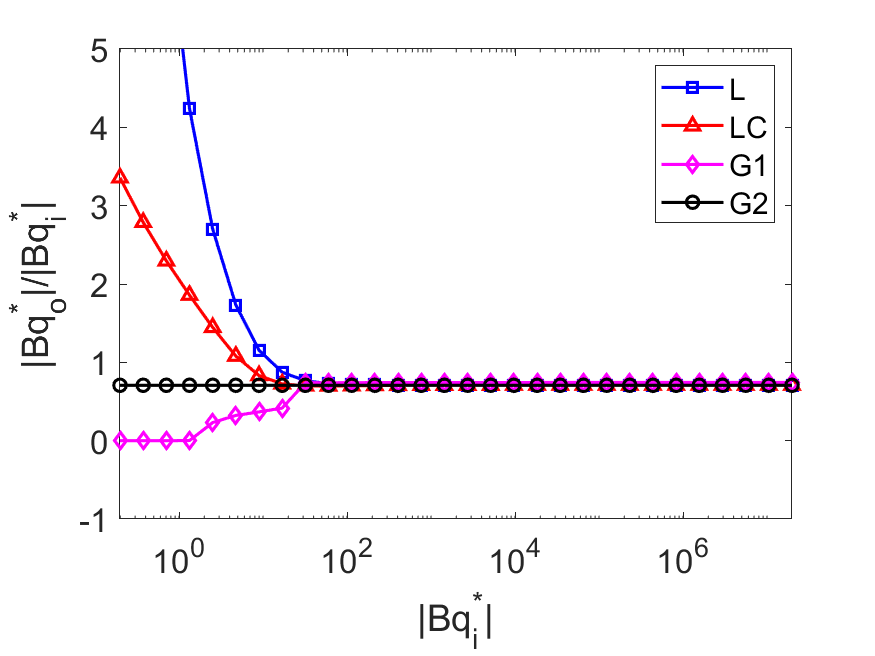}
  \end{minipage}
  \begin{minipage}{0.5\textwidth}
  \centering  \includegraphics[width=\linewidth]{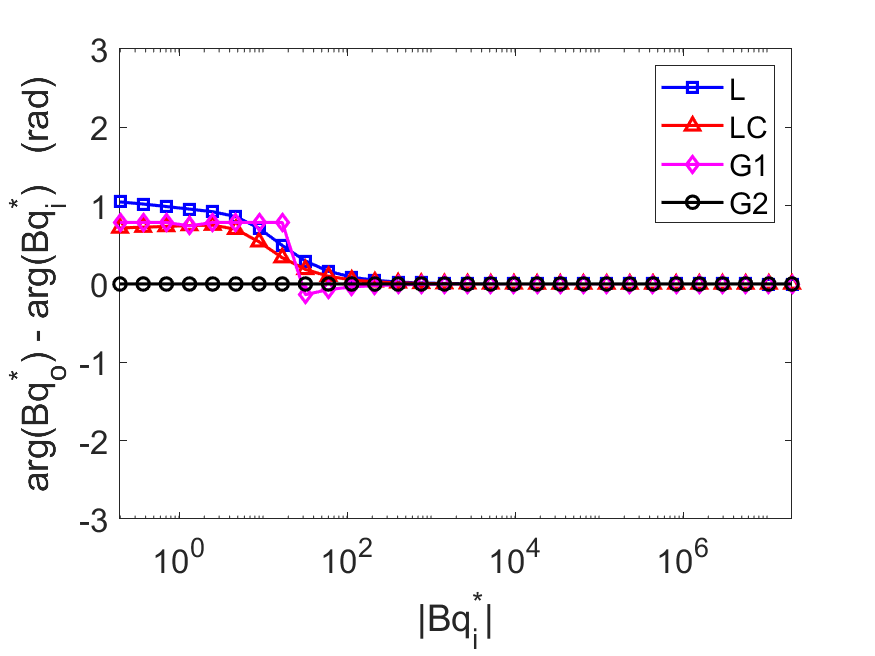}
  \end{minipage}
  \begin{minipage}{0.5\textwidth}
  \centering  \includegraphics[width=\linewidth]{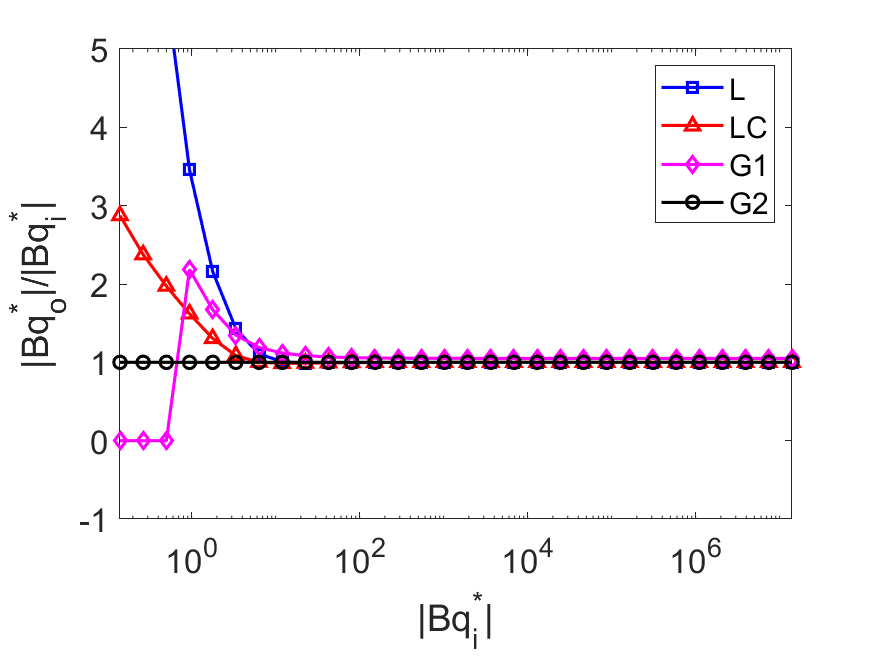}
  \end{minipage}
  \begin{minipage}{0.5\textwidth}
  \centering  \includegraphics[width=\linewidth]{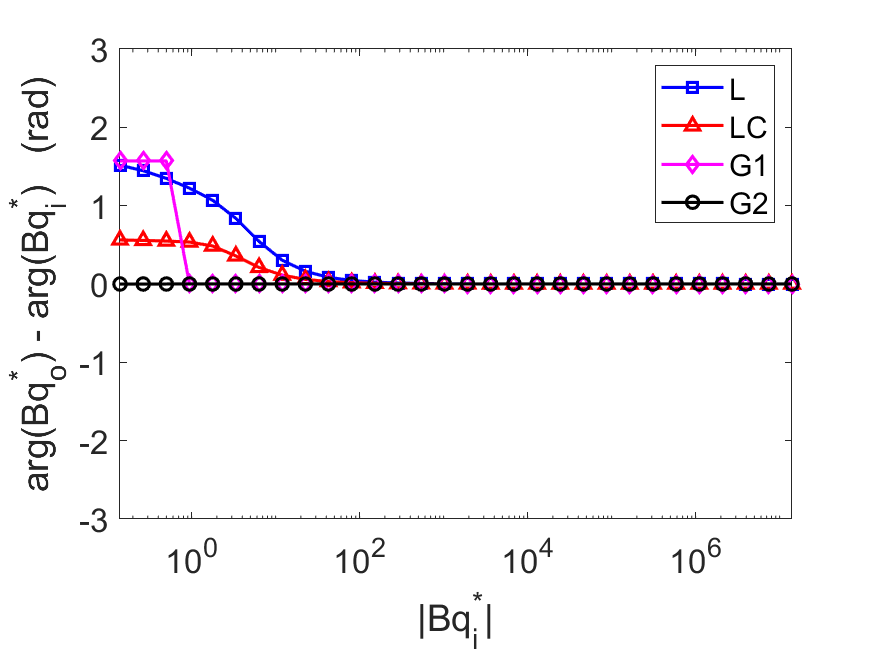}
  \end{minipage}
\caption{Comparative performance of four data processing schemes: the linear approximation with and without calibration compensation, and the G1 and G2 codes for air/water interfaces. Linear approximation: blue squares, linear approximation with calibration compensation: red triangles, G1 code: magenta diamonds, and FFBDA software package: black circles. Left and right panels: ratio of the moduli and argument difference between the output and input interfacial viscosities. Upper row panels: purely viscous interfaces sheared at $\omega = 10$ rad/s. Middle row panels: viscoelastic interfaces sheared at $\omega = 1$ rad/s. Lower row panels: purely elastic interfaces sheared at $\omega = 0.1$ rad/s.}
\label{fig:comp_perf0_omega}
\end{figure}

In Figure \ref{fig:comp_perf0_omega} left and right panels show, respectively, the ratio of the moduli of the input and output interfacial viscosities, $|Bq^*_o|/|Bq^*_i|$, and the difference between the corresponding arguments, $\arg (Bq^*_o) - \arg (Bq^*_i)$ for air/water interfaces. Consequently, $|Bq^*_o|/|Bq^*_i| = 1$, and $\arg (Bq^*_o) - \arg (Bq^*_i) = 0$ indicate that the input and output $Bq^*$ values coincide perfectly. Blue square symbols correspond to the simplest L scheme, red triangles to the LC scheme, magenta diamonds to the G1 code, and black circles to the G2 code here presented. The panels at the top, middle, and bottom rows correspond, respectively, to the cases of interfaces purely viscous ($Bq^* = Bq$) sheared at $\omega = 10$ rad/s, viscoelastic ($Bq^* = Bq(1 - i)$) sheared at $\omega = 1$ rad/s, and purely elastic ($Bq^* = -iBq$) sheared at $\omega = 0.1$ rad/s interfaces. 

The most salient feature of the comparison is the the G2 code performs extremely well, achieving a moduli ratio of 1 and an a null argument difference, for all values of $Bq^*$, whatever the mechanical character of the interfaces. On the other hand, the L, LC, and G1 packages all yield good results for large values of $Bq^*$, but for $Bq^*_i \le 10$, the L and LC schemes grossly overestimate the values of both the modulus and argument of $Bq^*$, while the G1 scheme clearly underestimates the value of $Bq^*$. The results here shown demonstrate that the ad-hoc iterative procedure defined for the G1 code does not yield the same fixed points provided by the iterative map defined for the G2 code at low values of $Bq^*$.

\section{Conclusions}

The present report describes and makes publicly available a second generation (G2) code for the analysis of experimental data obtained by means of any DWR rotational ISR. The code accepts both raw torque and inertia corrected torque input data. 

The G2 code incorporates the following improvements upon the first generation (G1) code are: i) a major improvement of the discretization scheme by allowing any even number of mesh nodes upon the ring's cross-section diagonal, hence allowing for a much improved spatial resolution, ii) using a second order centred finite differences scheme to calculate the gradients of the velocity fields close to the ring, iii)  the implementation of a physically based iterative scheme using the probe's equation of motion, and iv) a convergence criterion based on the experimentally obtained complex amplitude ratio.

We have illustrated the performance of the G2 code by, first, studying the calculated velocity fields at the bulk phases and the radial velocity profiles at the interface in a wide range of the input parameters, showing that the G2 code yields a much smoother interfacial velocity gradient than the G1 code. 

Second, we have tested the error rejection performance of the G2 code by consistency and error propagation tests that have allowed us to make well founded recommendations of the code parameters related with the mesh spacing, \texttt{ringSubs} $ = 40$, and the convergence tolerance, \texttt{tolMin} $= 10^{-5}$.

Finally, we have compared the results of the iterative obtaining of the complex Boussinesq number, $Bq^*_o$, which contains the complex viscosity, $\eta^*_s$, against the results obtained by two simpler methods that do not take into account the interface-bulk phases coupling and the G1 code that properly includes the interface-bulk phases coupling. The results show that while all of the methods give reasonable results for high values of $Bq^*$, only the G2 code yields correct values when $Bq^* \lesssim 100$.

\section*{Acknowledgments}
The authors gratefully acknowledge the support of the Spanish Ministerio de Ciencia e Innovación - Agencia Estatal de Investigación (MCIN/AEI/10.13039/501100011033) through
project PID2020-117080RB-C54, and fruitful discussions with J. Tajuelo, J.M. Pastor, M. Rodríguez-Hakim, P. Gutfreund, and A. Maestro. P.S.P. acknowledges the MICINN-ILL post-doc program for supporting his stay at ILL.

\newpage

\appendix

\section{Example of a starting script}
\label{section:exampleScript}

\begin{verbatim}
% This is an example of a script that calls postprocessing_DWR_ll.m
close all
clear,clc

% Geometry parameters
geom = struct();

% geom.small = struct();
% geom.small.H = 3/1000;
% geom.small.R6 = 18/1000;
% geom.small.R5 = 17/1000;
% geom.small.Rr = 17.5;
% geom.small.R1 = 13.5/1000;
% geom.small.R3 = 22.66666666666/1000;
% geom.small.ringW = geom.small.R6 - geom.small.R5;
% geom.small.stepW = geom.small.ringW;
% geom.small.R4 = geom.small.R3 + geom.small.stepW;
% geom.small.R2 = geom.small.R1 - geom.small.stepW;
% geom.small.G1 = geom.small.R5 - geom.small.R1; 
% geom.small.G2 = geom.small.R3 - geom.small.R6;
% geom.small.inertia = (0.09959920 + 0.0009)/1000; % system (rotor + DWR) 
                                                   inertia [Kg·m^2]

geom.medium = struct();
geom.medium.H = 3/1000;
geom.medium.R6 = 24.5/1000;
geom.medium.R5 = 23.5/1000;
geom.medium.Rr = 24;
geom.medium.R1 = 20/1000;
geom.medium.R3 = 28.7875/1000;
geom.medium.ringW = geom.medium.R6 - geom.medium.R5;
geom.medium.stepW = geom.medium.ringW;
geom.medium.R4 = geom.medium.R3 + geom.medium.stepW;
geom.medium.R2 = geom.medium.R1 - geom.medium.stepW;
geom.medium.G1 = geom.medium.R5 - geom.medium.R1; 
geom.medium.G2 = geom.medium.R3 - geom.medium.R6;
geom.medium.inertia = (0.09959920 + 0.00128880)/1000; % system (rotor + DWR) 
                                                      inertia [Kg·m^2]
geom.medium.ICorrected = false;

% Mesh parameters
mesh = struct();
mesh.ringSubs = 40;% Must be pair!!
if rem(mesh.ringSubs, 2) ~= 0
    warning('Ring subs must be pair!!')
end
mesh.upperBC = 'fb'; % free boundary 'fb' or no-slip 'ns'
mesh.DOrder = 2; % first (1) or second (2) order

% Upper phase physical parameters
Bulk = struct();
Bulk.rho_bulk1 = 1000;% kg/m^3 water
Bulk.eta_bulk1 = 1e-03;% Ns/m water
Bulk.rho_bulk2 = 1.204;% kg/m^3 air
Bulk.eta_bulk2 = 1.813e-05;% Ns/m air

% Iterative scheme parameters
iteParams = struct();
iteParams.iteMax = 100;% maximum number of iterations
iteParams.tolMin = 1e-5;% threshold tolerance

% Input/output data
IO = struct();
IO.colIndexFreq = 1; % ordinal number of the data of the column
                     that contains the modulus of the frequency
IO.colIndexAR = 2; % ordinal number of the data of the column
                     that contains the modulus of the amplitude ratio
IO.colIndexDelta = 3; % ordinal number of the data of the column
                      that contains the modulus of the phase shift
IO.colIndexAmp = 6; % ordinal number of the data of the column
                    that contains the angular amplitude
IO.colIndexTorq = 4; % ordinal number of the data of the column 
                     that contains the Torque amplitude
IO.colIndexRho = 8; % ordinal number of the data of the column
                      that contains the bulk density
IO.colIndexEta = 9; % ordinal number of the data of the column 
                      that contains the bulk viscosity
IO.inputFilepath = pwd; % input filepath
IO.outputFilepath = pwd; % output filepath

% Execute postprocessingDWR_ll.m with the specified input data
[resultStruct, timeElapsedTotal] = postprocessingDWR_ll(geom.medium, mesh,
Bulk, iteParams, IO);
save('MATLLABresults', 'resultStruct', 'timeElapsedTotal')
\end{verbatim}










\end{document}